\documentclass[a4paper,11pt]{article}
\pdfoutput=1
\usepackage{jcappub}
\usepackage{bm}
\usepackage{latexsym}
\usepackage{dcolumn}
\usepackage{amsfonts,amssymb,amsmath}
\usepackage{graphicx,epsfig}
\usepackage{psfrag}
\usepackage{subfigure}
\usepackage{rotating}
\usepackage{hyperref}
\usepackage{tikz}
\usepackage{soul}
\usepackage[utf8]{inputenc}

\hypersetup{
	bookmarks=true,         
	unicode=false,          
	pdftoolbar=true,        
	pdfmenubar=true,        
	pdffitwindow=false,     
	pdfstartview={FitH},    
	pdftitle={My title},    
	pdfauthor={Author},     
	pdfsubject={Subject},   
	pdfcreator={Creator},   
	pdfproducer={Producer}, 
	pdfkeywords={keyword1} {key2} {key3}, 
	pdfnewwindow=true,      
	colorlinks=true,       
	linkcolor=red,          
	citecolor=cyan,        
	filecolor=magenta,      
	urlcolor=green,           
	linktocpage=true
}

\def\beq{\begin{equation}}
\def\efq{\end{equation}}
\def\br{\begin{eqnarray}}
\def\er{\end{eqnarray}}
\def\benu{\begin{enumerate}}
	\def\efnu{\end{enumerate}}

\def\l{\left}
\def\r{\right}
\def\cR{{\cal R}}
\def\d{{\rm d}}
\def\f{\frac}

\def\cR{{\mathcal{R}}}
\def\cS{{\mathcal{S}}}

\def\Mpl{{M_{_{\textup{Pl}}}}}

\makeatletter
\gdef\@fpheader{}
\makeatother

\begin{document}
	\title{Generating PBHs and small-scale GWs in two-field models of inflation}%
	\author[1,2,3]{Matteo~Braglia,}
	\author[4,2,3]{Dhiraj~Kumar~Hazra,}
	\author[2,3]{Fabio~Finelli,} 
	\author[5,6,7,8]{George~F.~Smoot,}
	\author[9]{L.~Sriramkumar,}	 
	\author[10]{Alexei~A.~Starobinsky}
	\affiliation[1]{DIFA, Dipartimento di Fisica e Astronomia,\\
		Alma Mater Studiorum, Universit\`a degli Studi di Bologna,\\
		Via Gobetti, 93/2, I-40129 Bologna, Italy}
	\affiliation[2]{INAF/OAS Bologna, Osservatorio di Astrofisica e Scienza dello Spazio, \\
		Area della ricerca CNR-INAF, via Gobetti 101, I-40129 Bologna, Italy}
	\affiliation[3]{INFN, Sezione di Bologna, \\
		via Irnerio 46, I-40126 Bologna, Italy}
				\affiliation[4]{The Institute of Mathematical Sciences, HBNI, CIT Campus, Chennai 600113, India}
	\affiliation[5]{AstroParticule et Cosmologie (APC)/Paris Centre for Cosmological Physics, Universit\'e
		Paris Diderot, CNRS, CEA, Observatoire de Paris, Sorbonne Paris Cit\'e University, 10, rue Alice Domon et Leonie Duquet, 75205 Paris Cedex 13, France}
	\affiliation[6]{Institute for Advanced Study \& Physics Department, Hong Kong University of Science and Technology, Clear Water Bay, Kowloon, Hong Kong}
	\affiliation[7]{Physics Department and Lawrence Berkeley National Laboratory, University of California, Berkeley, CA 94720, USA}
	\affiliation[8]{Energetic Cosmos Laboratory, Nazarbayev University, Astana, Kazakhstan}
	\affiliation[9]{Department of Physics, Indian Institute of Technology Madras, 
		Chennai 600036, India}
	\affiliation[10]{Landau Institute for Theoretical Physics, Moscow 119334, Russia}
    
	\emailAdd{matteo.braglia2@unibo.it,  dhiraj@imsc.res.in, fabio.finelli@inaf.it, gfsmoot@lbl.gov, sriram@physics.iitm.ac.in, alstar@landau.ac.ru} 

	\abstract
	{Primordial black holes (PBHs) generated by gravitational collapse of large primordial overdensities can be a fraction of the observed dark matter.
	In this paper, we introduce a mechanism to produce a large peak in the primordial power spectrum (PPS) in two-field inflationary models characterized by two stages of inflation  based on a large non-canonical kinetic coupling. This mechanism is generic  to several two-field inflationary models,  due to a temporary tachyonic instability  of the isocurvature perturbations at the transition between the two stages of inflation. We numerically compute the primordial perturbations from largest scales to the small scales corresponding to that of PBHs using an extension of BINGO (BI-spectra and Non-Gaussianity Operator). Moreover we numerically compute the stochastic background of gravitational waves (SBGW) produced by second order scalar perturbations within frequencies ranging from nano-Hz to KHz that covers the observational scales corresponding to Pulsar Timing Arrays, Square Kilometer Array to that of Einstein telescope. We discuss the prospect of its detection by these proposed and upcoming gravitational waves experiments.}
	\maketitle
	
	
	\section{Introduction}
	
Primordial Black Holes (PBHs), black holes formed in the Early Universe, before Big Bang nucleosynthesis~\cite{Bertone:2018xtm}, could explain a fraction of the observed cold dark matter (CDM) abundance.
The CDM fraction in the form of PBHs is tightly constrained by current cosmological and astrophysical observations and in the ideal case of a monochromatic distribution, only PBHs of masses  around $10^{-12}$ and $10^{-16} M_\odot$ can account for the total amount of the CDM. However, the recent detection of $\sim 30 M_\odot$ black holes coalescence~\cite{Abbott:2016blz} has renewed the interest for PBHs~\cite{Bird:2016dcv,Clesse:2016vqa,Sasaki:2016jop} and it has also been realized that a small PBH-to-CDM fraction could still be phenomenologically important.

Possible mechanisms of PBHs formation include domain walls, vacuum bubbles nucleation and cosmic string loops (see~\cite{Sasaki:2018dmp} and references therein). However, the standard mechanism to produce PBHs within the inflationary scenario is the gravitational collapse during radiation era of large small-scale overdensities produced during the last stages of inflation, close to its end.

A variety of models that produce a large peak in the curvature power spectrum during inflation exist in the literature. This can be achieved in single-field inflation scenarios by using a local feature in the inflaton potential like a rapid change of its amplitude or a break in its first derivative~\cite{Starobinsky:1992ts,Ivanov:1994pa,Motohashi:2019rhu}, an inflection point~\cite{Garcia-Bellido:2017mdw,Motohashi:2017kbs,Germani:2017bcs,Ballesteros:2017fsr,Cicoli:2018asa,Byrnes:2018txb,Passaglia:2018ixg,Dalianis:2018frf,Bhaumik:2019tvl} or a tiny bump superimposed on it~\cite{Ozsoy:2018flq,Mishra:2019pzq}, non-trivial inflaton sound speed~\cite{Kamenshchik:2018sig,Cai:2018tuh,Ballesteros:2018wlw}\footnote{See also Refs.~\cite{Pattison:2017mbe,Biagetti:2018pjj,Ezquiaga:2018gbw} for the effects of quantum diffusion in the PBH abundance.}. More possibilities arise in multiple-field inflationary models, where large peaks in the power spectrum leading finally to PBHs can be generated both with curvature (adiabatic) perturbations~\cite{Polarski:1992dq,Starobinsky:1994mh} and  isocurvature ones~\cite{Kofman:1986wm,Kofman:1988xg,Polarski:1994rz,Starobinsky:1994mh,Starobinsky:2001xq}. Among these we mention hybrid inflationary models, where a second waterfall field triggers the growth of fluctuations near to the end of inflation~\cite{GarciaBellido:1996qt,Kawasaki:2006zv,Kawaguchi:2007fz,Frampton:2010sw,Clesse:2015wea}, models with couplings of the inflaton to scalaron fields~\cite{Starobinsky:2001xq,Pi:2017gih,Cheong:2019vzl} or with explosive production of gauge fields~\cite{Linde:2012bt,Garcia-Bellido:2016dkw,Domcke:2017fix}. 

In this paper, we build on our previous work~\cite{Braglia:2020fms} and study the generation of PBHs in a two-field model consisting of a canonical scalar field (say, $\phi$) and a second scalar field (say, $\chi$) with a non-canonical kinetic term of  the  form $f(\phi)(\partial\chi)^2$. As a specific example of this general mechanism, we choose a setting in which $\phi$ $(\chi)$ is the effectively heavier (lighter) field driving the first (second) stage  of  inflation, highlighting the role of the non-canonical coupling between the two scalar fields in sourcing a bump in the curvature power spectrum. As shown in Ref.~\cite{Braglia:2020fms}, when inflation consists of two stages, the effective mass of isocurvature perturbations can become temporarily negative around the transition from the first to the second stage of inflation if the non-canonical kinetic term is appropriately chosen.  Therefore, a temporary tachyonic growth of isocurvature perturbations enhances the curvature fluctuations, resulting in a bump in their primordial power spectra (PPS) at the scales that cross the Hubble radius around transition, as necessary to produce PBHs. Depending on the duration of the second stage of inflation, the bump occurs at different scales and PBHs of different masses can form.

Such a bump in the PPS, however, does not only lead to the production of PBHs. In fact, large scalar overdensities act as a source for a stochastic background of gravitational waves (SBGW) at the second order in perturbation theory~\cite{Matarrese:1993zf,Matarrese:1997ay,Noh:2004bc,Carbone:2004iv}. The peak of such GWs is related to the mass of the produced PBHs and, in turn, to the scale at which the PPS peaks. For example, the collapse of large scalar fluctuations at scale $k\sim 10^{12}\, \text{Mpc}^{-1}$ into  $M_{\textup{PBH}}\sim 10^{-12}\,M_\odot$ PBHs leads to a SBGW that peaks in the frequency band targeted by the future space based GW interferometer  LISA~\cite{Wang:2016ana,Bartolo:2018evs}.   
According to these results, the duration of the second stage of inflation becomes critical in determining how the resulting SBGW falls into the sensitivity range of different forthcoming experiments.
Note that the production of SBGW is independent of whether PBHs are formed or not.  In fact, a tiny decrease in the peak in the PPS can drastically affect the PBHs mass fraction, without sizeably changing the SBGW predictions.  

This paper is organized as follows. In the following section, we  introduce our toy model and analyze its background evolution, discussing the main features that will lead to the growth of scalar perturbations.
In~\autoref{sec:spectra},  we present the results of our numerical computation for the PPS of scalar and tensor perturbations. In order to fully take into account the coupling between curvature and isocurvature perturbations, we resort to a  numerical computation. From the scalar PPS, we compute the PBH mass fraction in~\autoref{sec:PBH} and   the resulting SBGW in~\autoref{sec:GW} and comment on the role of the different parameters at play. In~\autoref{sec:coupling}, we comment on the consequences on the PBHs mass fraction and SBGW signal of choosing different functional forms for the non-canonical coupling between the two fields.
	We discuss our results in the concluding~\autoref{sec:discussion}. In~\autoref{appendix:VaryingMass}, for completeness, we provide an additional scan of the parameter space for the model presented in~\autoref{sec:spectra} and we collect some analytical results about the background behavior of our model in~\autoref{appendix:Background}.
	All the results presented in this paper are obtained with a two-field extension  of the code BINGO~\cite{Hazra:2012yn}, also used in Ref.~\cite{Braglia:2020fms}, which we have further modified to compute the mass fraction of PBHs and the 
	relic density of GWs.

	
	\section{Theoretical construction}~\label{sec:background}
	
	In this section, we  introduce our  two-field toy model of inflation.
    We do not present the equations governing  the 
	perturbations in this paper. 
	We would refer the reader to our earlier paper~\cite{Braglia:2020fms} 
	as well as the original efforts for further details in this 
	regard~\cite{Gordon:2000hv,Starobinsky:2001xq,DiMarco:2002eb,
	DiMarco:2005nq,Lalak:2007vi}.
	We would also refer the reader to~\autoref{appendix:Background} for the  equations and some analytical expressions for the background to better understand our numerical results.
	
	The dynamics of the two scalar fields is governed by the following 
	action
	\begin{equation}
	\label{action}
	S[\phi,\chi]=\int \d^4x\,\sqrt{-g}\,
	\left[\frac{\Mpl^2}{2} R
	-\frac{1}{2}(\partial \phi)^2
	-\frac{f(\phi)}{2}(\partial\chi)^2-V(\phi,\chi)\right]
	\end{equation}
	and 	we  work with the spatially flat 
	Friedmann-Lema{\^i}tre-Robertson-Walker 
	(FLRW) universe described by the line-element 
	\begin{equation}
	\d s^2=-\d t^2+a^2(t)\,\d{\bm x}^2,\label{flrw} 
	\end{equation}
	where $a(t)$ is the scale factor and $t$ is the cosmic time.
	Note that, while $\phi$ is a canonical scalar field, $\chi$ is 
	a non-canonical scalar field due to the presence of the function
	$f(\phi)$ in the term describing its kinetic energy.
	Evidently, apart from the potential $V(\phi,\chi)$, through which 
	the fields can in principle interact, the function $f(\phi)$ also leads to an
	interaction between the fields. In order to connect with the equations of Refs.~\cite{DiMarco:2002eb,
		DiMarco:2005nq} we define $f(\phi)\equiv{\rm e}^{2 b(\phi)}$.
	
		As a toy model, we consider the following decoupled potential where the $\phi$ has the KKLTI form~\cite{Kallosh:2018zsi} and $\chi$ has a simple quadratic potential:
	
	\begin{equation}
	\label{eq:potential}
	V(\phi,\,\chi)=V_0\frac{\phi^2}{\phi_0^2+\phi^2}+\frac{m^2_\chi}{2}\chi^2.
	\end{equation}
	Note that this is the same used in Ref.~\cite{Braglia:2020fms} with $\phi\leftrightarrow\chi$. We stress that the potential used in~\autoref{eq:potential} is only a toy model and the results of this paper do not rely on this particular realization. We  consider two different non-canonical couplings in the rest of the paper:
	\begin{align}
	\label{eq:coupling1}
	&f_A(\phi)=e^{2 b_A(\phi)}\equiv e^{2 b_1\phi},\\
	\label{eq:coupling2}
	&f_B(\phi)=e^{2 b_B(\phi)}\equiv e^{2 b_2\phi^2}.
	\end{align}
	
		\begin{figure}
		\begin{center} 
			\resizebox{214pt}{172pt}{\includegraphics{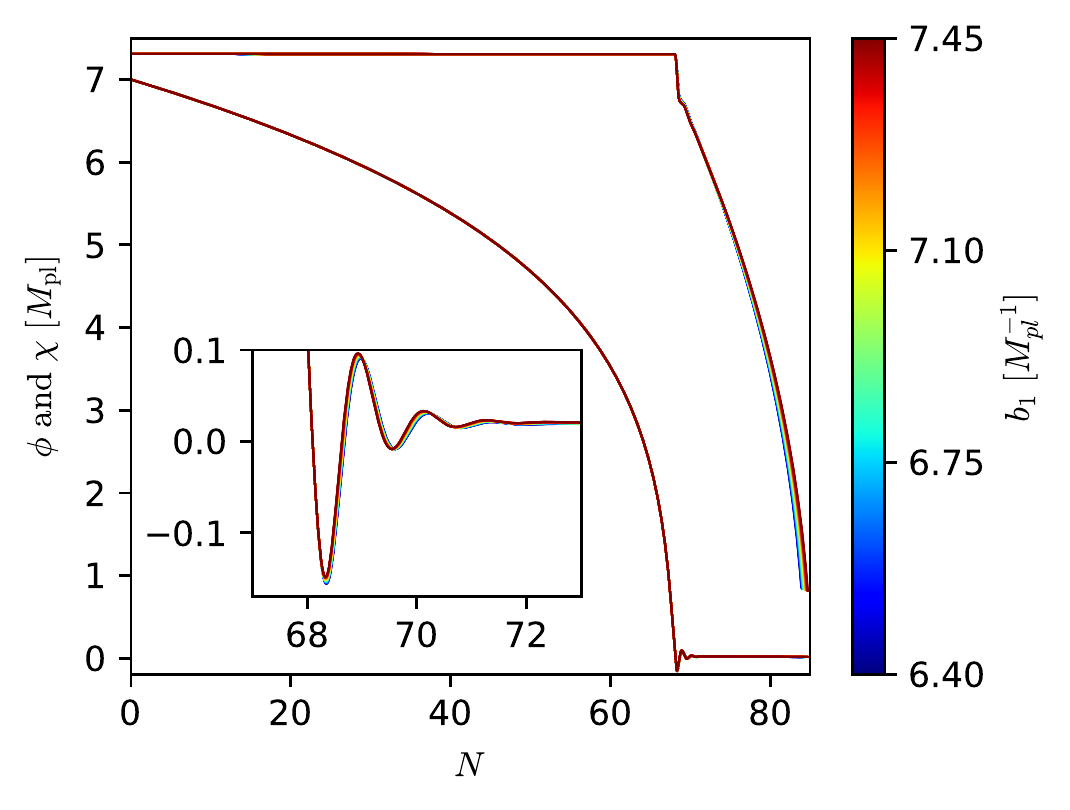}}
			\resizebox{214pt}{172pt}{\includegraphics{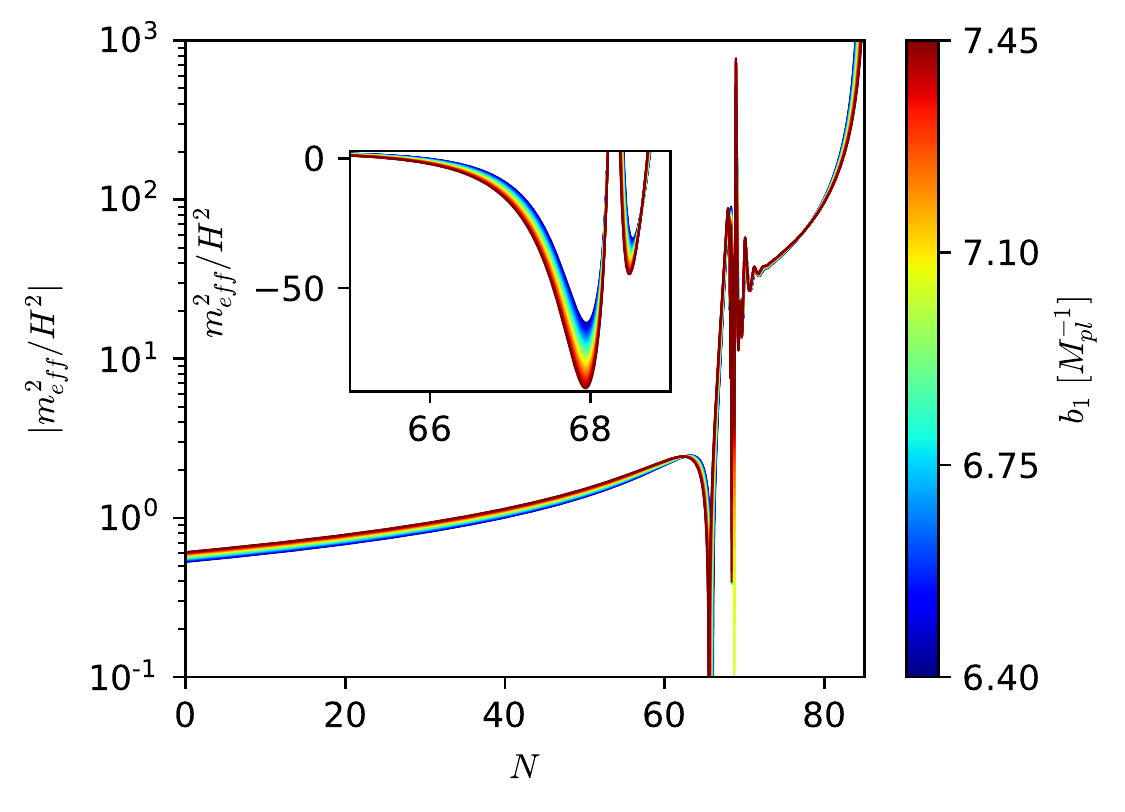}}	
			\resizebox{214pt}{172pt}{\includegraphics{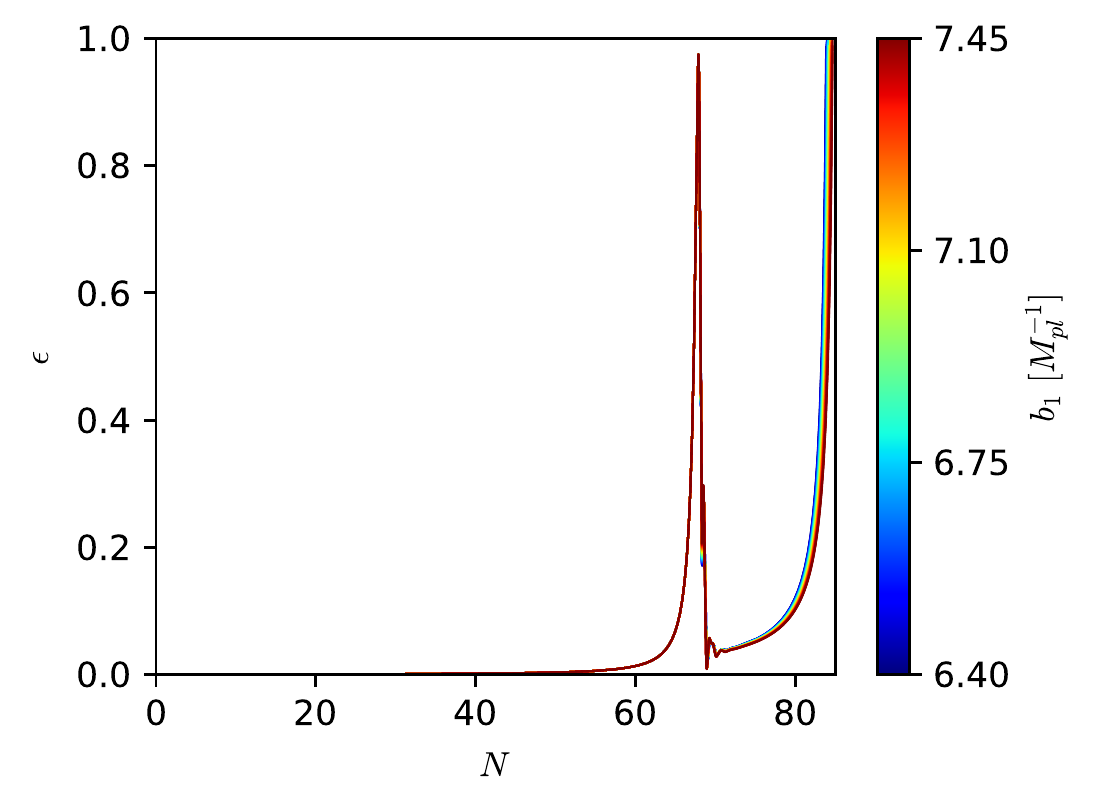}}
			\resizebox{214pt}{172pt}{\includegraphics{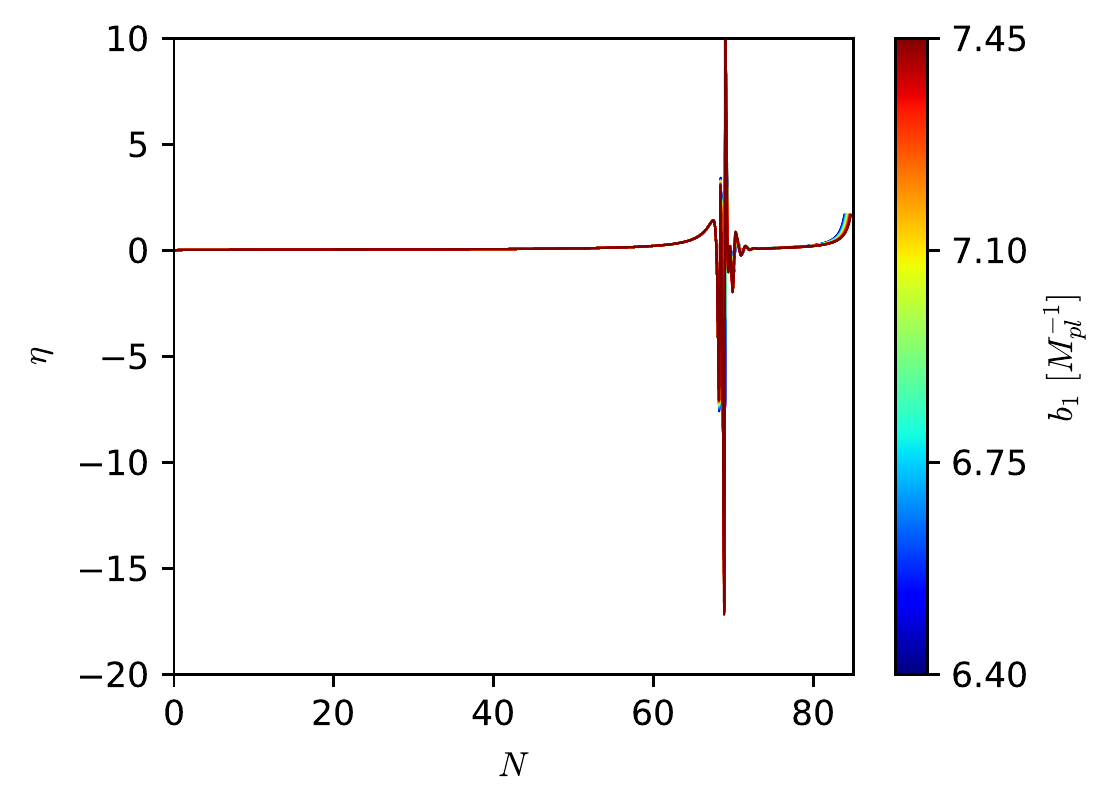}}	
		\end{center}
		\caption{\label{fig:Background} 
		[Top] Evolution of the scalar fields (left) and the effective mass of isocurvature perturbations $m_\textup{eff}^2$ as defined in~\autoref{eq:effmassiso}. [Bottom] Evolution of the first two slow-roll parameters $\epsilon$ (left) and $\epsilon_2\equiv\eta$ (right). The parameters used are the ones for the LISA case  are provided in the text, while $b_1$ is varied for a continuous range of values.}
	\end{figure}
	
	We define the Hubble Flow Functions (HFFs) or slow-roll parameters as 
	follows~\cite{Schwarz:2001vv}:
	\begin{equation}
	\epsilon_{i+1}\equiv\frac{\d\epsilon_i}{\d N},
	\end{equation}
	with
	\begin{equation}
	\epsilon_0\equiv\frac{H_{\textup{\rm in}}}{H}
	\end{equation}
	where $H_{\rm in}$ is the value of the Hubble parameter at some
	initial time during inflation, and $N=\int \d t\, H$ represents 
	the number of e-folds.
	We  refer to the regime wherein all the Hubble flow functions 
	are small (i.e. $\epsilon_i\ll1$, for all $i>0$) as the slow-roll 
	regime.

	In ~\autoref{fig:Background}, we plot the relevant background quantities 
	for the illustrative case of $\phi_0=\sqrt{6}M_\textup{pl}$,  $V_0/(m_\chi\,M_\textup{pl})^2=500$ and $V_0$ fixed to  produce the correct COBE normalization at CMB scales (see next section). For definiteness, we use $f_A(\phi)$ and vary $b_1$ in order to highlight the effects of the coupling with the $\chi$ kinetic term. 
We choose $\phi_i=7.0 M_\textup{pl}$ and $\chi_i=7.31 M_\textup{pl}$ for the initial values of the scalar fields and we fix their initial time derivatives by imposing slow-roll initial conditions on $\dot{\phi}_i$ and $\dot{\chi}_i$. We   discuss sets of parameters that are relevant for observations in the next section.    

	 As can be easily seen from~\autoref{fig:Background}, the heavier of the two fields, i.e. $\phi$, rolls down its potential driving a first phase of inflation  while the lighter field $\chi$ remains frozen. 
	 When the first stage of inflation dominated by $\phi$  finishes, $\phi$ undergoes a few damped oscillations around its effective minimum and  the field $\chi$ starts a second inflationary phase that lasts around $\sim 20$ {\it e}-folds.  In the central panels of~\autoref{fig:Background}, the first slow-roll parameter $\epsilon_1\equiv\epsilon$ shows a bump between the two phases and the slow-roll conditions are violated, i.e. $\epsilon_2\equiv\eta>1$.

With this choice of parameters, the non-canonical kinetic term affects the 
the isocurvature mass (see~\autoref{eq:effmassiso} for its definition and next section for a discussion) that becomes temporarily negative at the transition between the first and the second stage of inflation, as  shown in~\autoref{fig:Background}. This plays an important role in the production of PBHs and GWs.

\section{Generating features in the primordial power spectrum}\label{sec:spectra}
We now compute numerically the scalar and tensor power spectra at the end of inflation. We evolve the curvature and isocurvature perturbation defined as:
\begin{equation}
\mathcal{R}= \frac{H}{\dot{\sigma}}Q_\sigma,\quad
\mathcal{S}=\frac{H}{\dot{\sigma}} Q_s.
\end{equation}
$Q_\sigma$ and $Q_s$ are the Mukhanov-Sasaki variables associated to the perturbations that are parallel and orthogonal to the field-space trajectory, i.e.~\cite{DiMarco:2002eb}
\begin{align}
\label{adi}
\delta\sigma=&\cos\theta\, \delta\phi+\sin\theta\, e^b\, \delta\chi,\\
\label{iso}
\delta s=&-\sin\theta\,\delta\phi+\cos\theta\, e^b\, \delta\chi,
\end{align}
where $\cos\theta\equiv  \dot{\phi}/\dot{\sigma}$, $\sin\theta\equiv e^b \dot{\chi}/\dot{\sigma}$ and $\dot{\sigma}=\sqrt{\dot{\phi}^2+e^{2 b}\dot{\chi}^2}$. 
They are given by $Q_\sigma=\delta \sigma+(\dot{\sigma/H})\,\Phi$, where $\Phi$ is the Newtonian gauge potential,
and $Q_s=\delta s$ and we should point out that $\delta s$ is an intrinsically gauge-invariant quantity.

The curvature and isocuvature power and cross power spectra are given by:
\begin{subequations}
	\label{P1P2}
	\begin{eqnarray}
	\mathcal{P}_{\cR}(k)
	&=&\frac{k^3}{2\pi^2}
	\l(\lvert\cR_1\rvert^2+\lvert\cR_2\rvert^2\r)
	=\mathcal{P}_{\cR_1}(k)+\mathcal{P}_{\cR_2}(k),
	\label{eq:PR}\\
	\mathcal{P}_{\cS}(k)
	&=&\frac{k^3}{2\pi^2}
	\l(\lvert\cS_1\rvert^2+\lvert\cS_2\rvert^2\r),\\
	\mathcal{C}_{\cR\cS}(k)
	&=&\frac{k^3}{2\pi^2}
	\l(\cR^\ast_1\cS_1+\cR^\ast_2\cS_2\r).
	\end{eqnarray}
\end{subequations}
where the subscript $1$ denotes the set of solutions integrated by imposing the Bunch-Davies initial conditions on $Q_\sigma$ and 
assuming the initial value of $\delta s$ to be zero, whereas the subscript $2$
implies vice versa. We evaluate all the spectra  at the end of inflation. For the full set of equations governing the dynamics of $\mathcal{R}$ and $\mathcal{S}$, and a more detailed explanation of the numerical procedure, we refer the interested reader to  Ref.~\cite{Braglia:2020fms}. As in Ref.~\cite{Braglia:2020fms}, we normalize the scale factor $a(N)$ so that the pivot scale $k_*=0.05$ $\text{Mpc}^{-1}$ crosses the Hubble radius $N_*=50$ $e$-folds before the end of inflation.

\begin{figure}
	\begin{center} \resizebox{453pt}{170pt}{\includegraphics{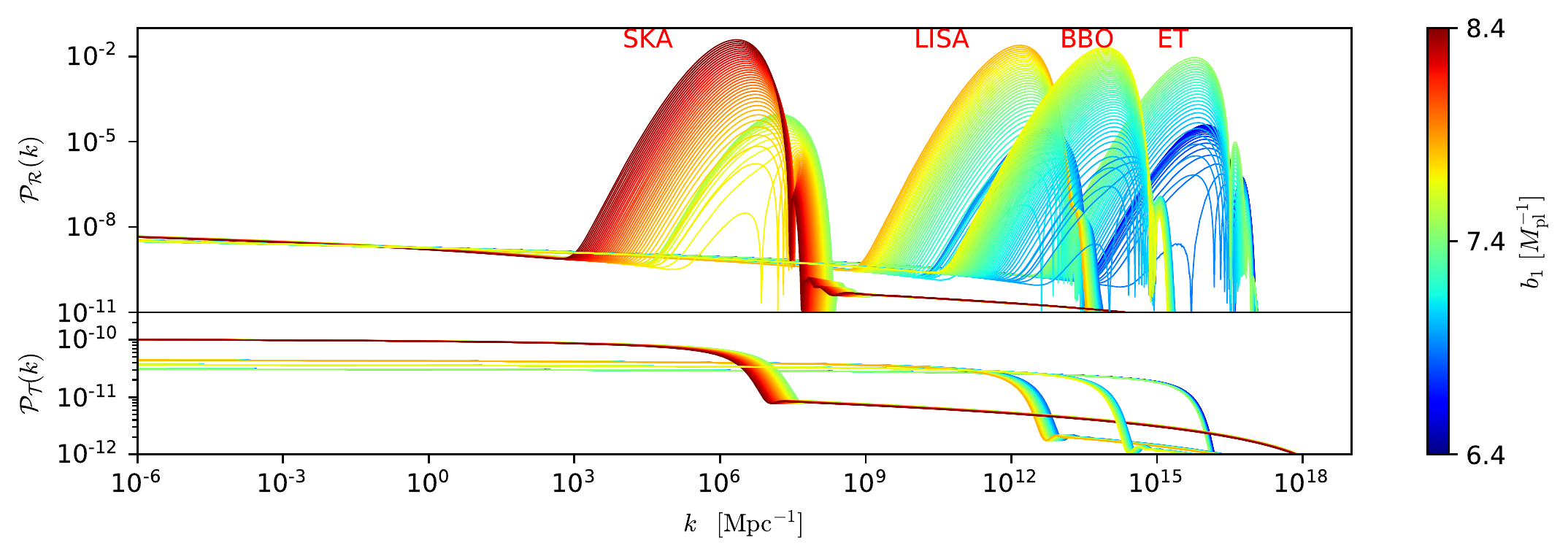}}
		\caption{\label{fig:power3} [Top] scalar and [bottom]  tensor power spectra at the end of inflation. The parameters used are given in the Table in the main text  and  $b_1$ is varied over a continuous range of values.}
		\end{center} 
\end{figure}

We present the results of our analysis in~\autoref{fig:power3}, where we have plotted four different examples\footnote{The name of each  example is chosen according to the frequency regime of  their associated SBGW (see next section). SKA, LISA, BBO and ET stand for Square Kilometer Array, Laser Interferometer Space Antenna, Big Bang Observer and Einstein Telescope respectively.} using the same potential parameters used in the last section and changed $V_0$ to  produce the correct COBE normalization for each case.  

All the power spectra consist of a nearly scale-invariant part at large scales and at very small that cross the Hubble radius during the first and the second stage of inflation respectively and a bump at the scales that cross the Hubble radius during the transition between the two stages. Depending on the duration of the second stage of inflation, which in turn depends on the initial condition on the lighter field $\chi$, the peak in the power spectrum changes location and the predictions at CMB scales change.

The initial field values for each case, together with the spectral  indices $n_s$ and the tensor-to-scalar ratio at the scale $k=0.002 \,\text{Mpc}^{-1}$, are given in Table \ref{tab1}.

\begin{table*}\begin{tabular}{ |p{2cm}||p{2cm}|p{2cm}  |p{2cm}  |p{2cm}  |}
	\hline
	&  $\phi_i \,[M_\textup{pl}]$&$\chi_i \,[M_\textup{pl}]$&$n_s$ & $r$ \\
	\hline
	SKA  & 7.0 &9.3 &0.9184 & 0.042\\
	LISA&  7.0  & 7.31&0.9537 &0.020 \\ 
	BBO & 7.0 & 6.55& 0.9601& 0.017\\
	ET& 7.0 & 5.6& 0.9640&0.014 \\
	\hline 
\end{tabular}\caption{\label{tab1} 
		Initial conditions on the scalar fields $\phi$ and $\chi$ and spectral index and tensor-to-scalar  for the spectra in Fig.~\ref{fig:power3}.}
\end{table*}
\bigbreak

We note that the spectral index becomes more red as $\chi$ increases and the peak in the power spectrum moves to larger scales. Indeed, for scales that cross the Hubble radius far from the transition the prediction are essentially those of single field inflation Ref.~\cite{Braglia:2020fms}. Therefore, as the second stage of inflation gets longer,  CMB scales cross the horizon when the inflaton $\phi$ is in a less flat region of its potential and the spectral index gets redder.
 
In particular, note that the SKA  example is in  tension  with the constraints on $n_s$ from current cosmological CMB data~\cite{Akrami:2018odb}.  However, we emphasise that  choosing a different form for $V(\phi)$ that gives a bluer power spectrum can improve the agreement of the SKA example with CMB constraints.

\begin{figure}
		\begin{center} 
			\resizebox{143pt}{115pt}{\includegraphics{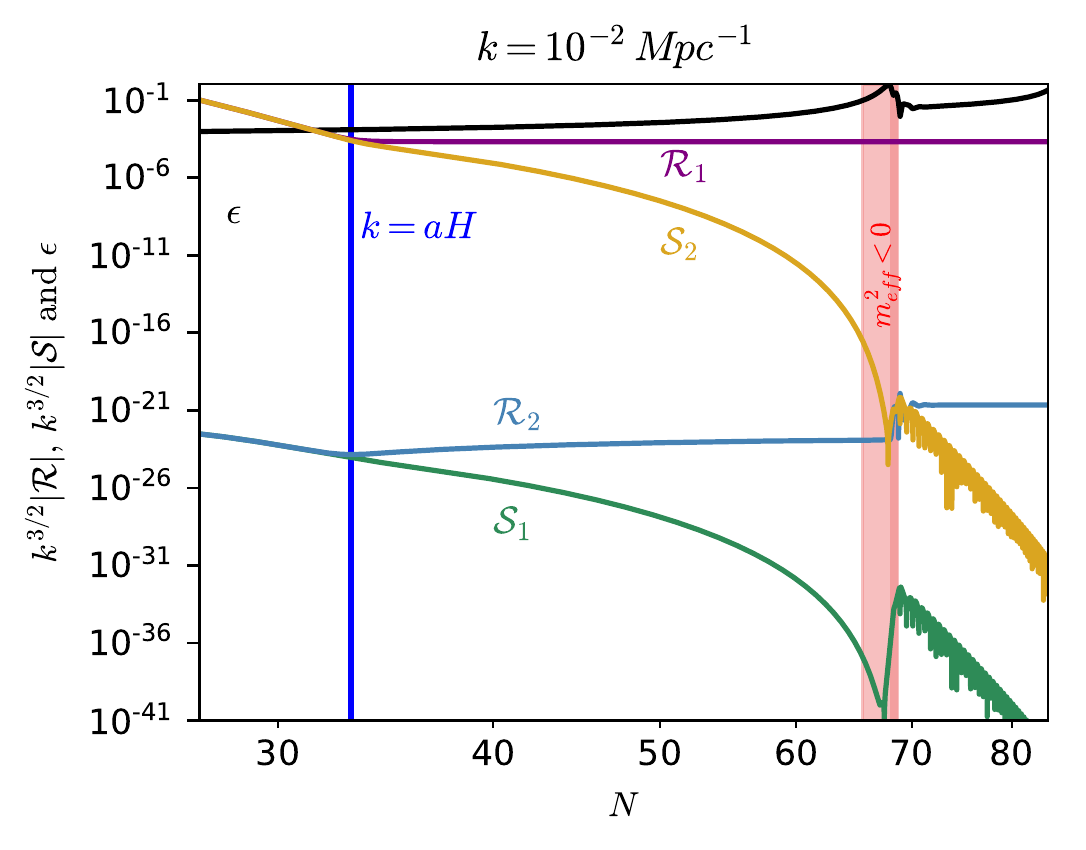}}		\resizebox{143pt}{115pt}{\includegraphics{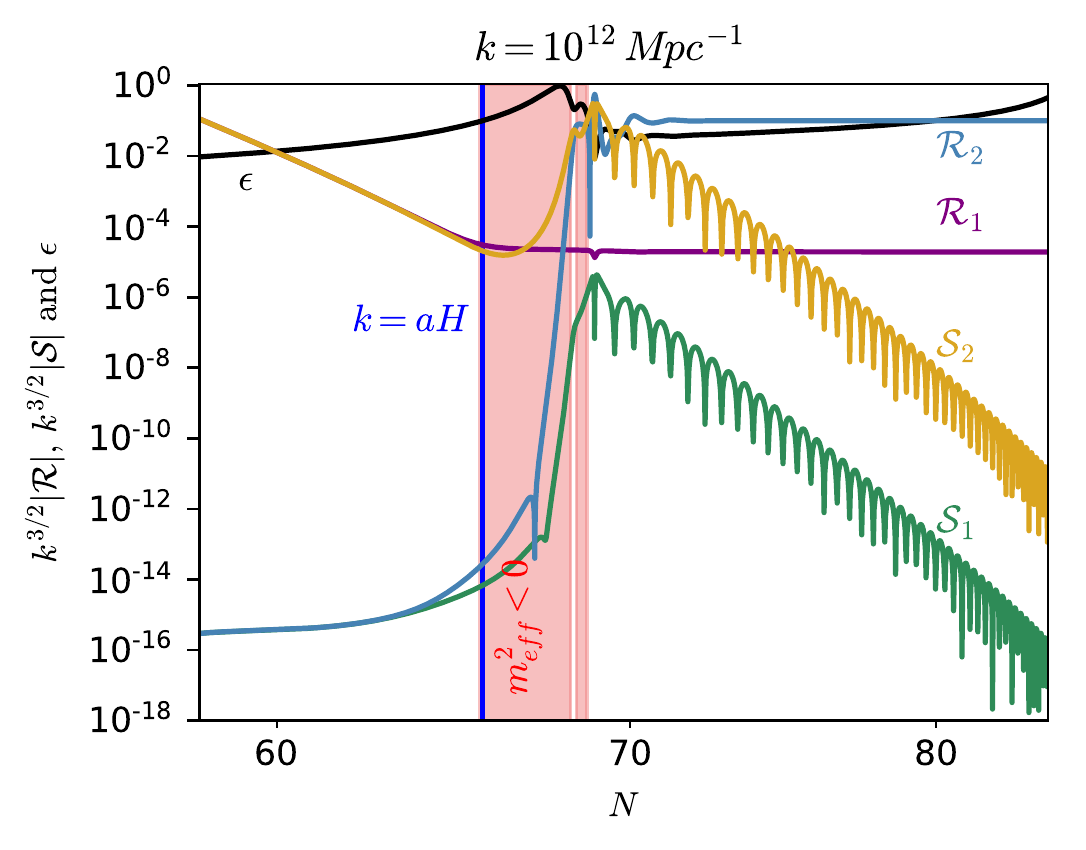}}		\resizebox{143pt}{115pt}{\includegraphics{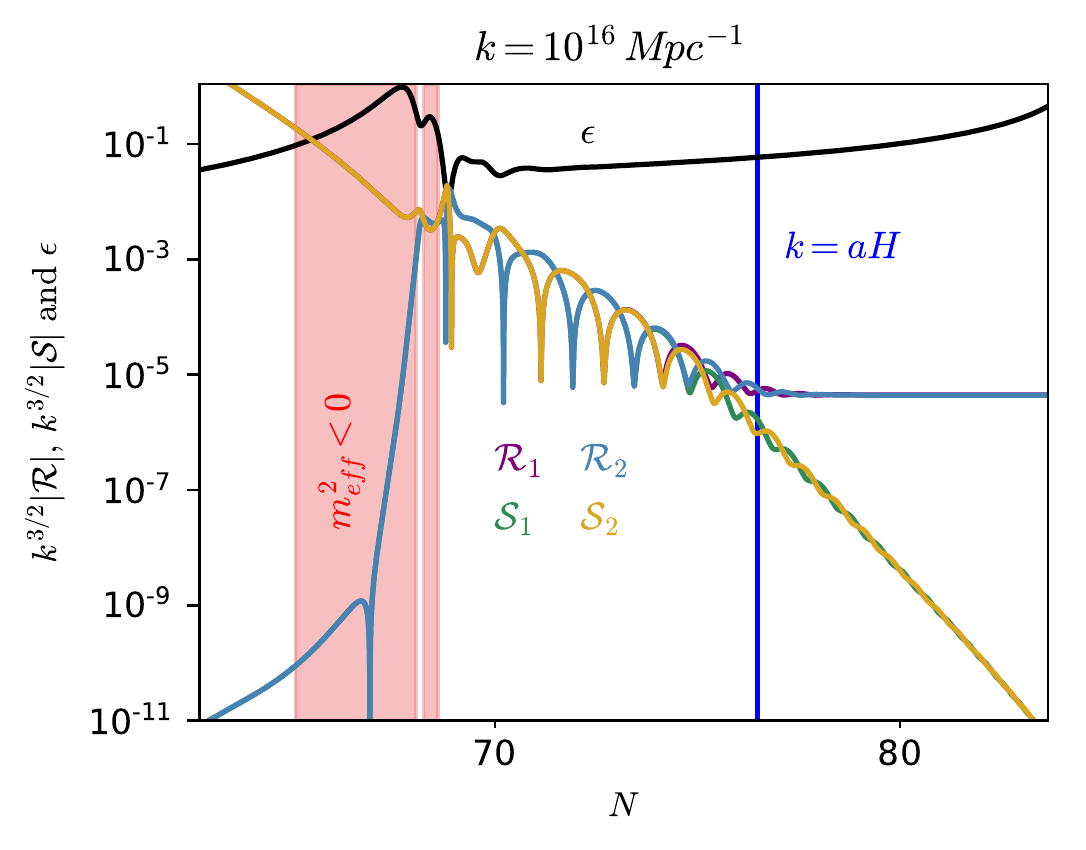}}	
		\end{center}
		\caption{\label{fig:modeEvolution}Evolution of the perturbed modes $k^{3/2}\lvert \mathcal{R}\rvert$ and $k^{3/2}\lvert \mathcal{S}\rvert$ for the representative LISA case. We plot the evolution  for the modes with $k_L= 10^{-2}\,\textup{Mpc}^{-1}$ [Left], $k_I= 10^{12}\,\textup{Mpc}^{-1}$ [Center] and $k_S= 10^{16}\,\textup{Mpc}^{-1}$ [Right]. The  blue vertical lines signal the $e$-folds when the modes cross the Hubble radius. We also plot the $\epsilon$ parameter in black lines. The light red shaded region is the one where $m_\textup{eff}^2<0$.} 
	\end{figure}

The crucial finding of this work is the large bump in the power spectra at small scales. As shown in Ref.~\cite{Braglia:2020fms}, when the coupling $f_1(\phi)$ is large enough, the isocurvature mass defined as

 \begin{equation}
 	\label{eq:effmassiso}
m_\textup{eff}^2\equiv V_{ss}+3\dot{\theta}^2
+b^2_\phi g(t)+b_\phi f(t)
-b_{\phi\phi}\dot{\sigma}^2
-4\f{V_s^2}{\dot{\sigma}^2},
\end{equation}
where, according to Ref.~\cite{DiMarco:2002eb},
\begin{subequations}
	\begin{eqnarray}
	g(t)&=&-\dot{\sigma}^2(1+3\sin^2\theta),\\
	f(t)&=&V_\phi(1+\sin^2\theta)-4 V_s\sin\theta,\\
	V_s&=&-V_\phi\sin\theta+{\rm e}^{-b}\,V_\chi\cos\theta,\\
			V_{ss} &=&V_{\phi\phi}\sin^2\theta
		- {\rm e}^{-b} V_{\phi\chi}\sin 2 \theta
		+ {\rm e}^{-2 b} V_{\chi\chi}\cos^2\theta ,\\
		V_{\sigma s} &=&-V_{\phi\phi}\cos\theta\sin\theta
		+ {\rm e}^{-b} V_{\phi\chi} (\cos^2\theta-\sin^2\theta)
		+ {\rm e}^{-2 b} V_{\chi\chi}\cos\theta\sin\theta,
	\end{eqnarray}
\end{subequations}
becomes temporarily negative at the transition between the two stages of inflation and leads to a transient tachyonic amplification of the isocurvature perturbations\footnote{See also Refs.~\cite{Renaux-Petel:2015mga,Brown:2017osf,Garcia-Saenz:2018ifx,Cicoli:2018ccr,Cicoli:2019ulk,Bjorkmo:2019aev,Bjorkmo:2019fls,Bjorkmo:2019qno,Chakraborty:2019dfh,Fumagalli:2019noh} for models of multi-field inflation where a temporary tachyonic instability of isocurvature pertburbations is induced by non-canonical kinetic terms.} and the sourcing to the curvature perturbation is more efficient, leading to a larger peak in $\mathcal{P}_\mathcal{R}$.  The tachyonic growth of isocurvature perturbations and the feedback to the curvature perturbations can also be appreciated from~\autoref{fig:modeEvolution}, where we plot the evolution of three perturbed modes in the representative LISA case, for the scales of $k_L=10^{-2}\,~\text{Mpc}^{-1}$, $k_B=10^{12}\,~\text{Mpc}^{-1}$ and $k_S=10^{16}\,~\text{Mpc}^{-1}$. As it is easy to see, isocurvature modes are  amplified when $m_\textup{eff}^2$ becomes negative. Even though this happens for all the modes in~\autoref{fig:modeEvolution}, only for the central plot this results in an effective amplification of the curvature perturbations. In fact, in the left panel, the isocurvature growth occurs much after Hubble crossing and it is not important, since the isocurvature modes have already decayed. In the opposite case, i.e. $k_S$,  the amplification occurs when the mode is still inside the Hubble radius and there is not any amplification of the curvature perturbation. The only region where the PPS is enhanced is thus the one of the scales that cross the Hubble radius during slow-roll violation.

Note that, despite the tachyonic amplification, super-horizon isocurvature modes soon decay after the transition and their power spectrum at the end of inflation is therefore very small.

The crucial result in~\autoref{fig:power3} is that the amplitude of the peak  can easily be  of the order $\mathcal{P}_\mathcal{R}(k_{\textup{peak}})\sim \mathcal{O}(0.01)$. This will be important in \autoref{sec:PBH} and \ref{sec:GW}, when we will consider the phenomenology of PBH formation and SBGW. 

Note that the spectral shape of the power spectra is very similar  in all the four examples  in~\autoref{fig:power3}. In particular, we note that tha maximal rate of growth is $n_s=4$, in agreement with the generic causality condition for
local physical processes producing inhomogeneities \cite{Abbott:1985di}. We show in~\autoref{appendix:VaryingMass} that a distinct shape can be obtained by varying the potential ratio. 

For completeness, we also plot the spectrum of tensor perturbations $\mathcal{P}_\mathcal{T}(k)$.  As can be seen from the Table above, our toy model produces a tensor power spectrum with $r\lesssim0.065$, as the KKLTI predictions for $r$ from the first stage of inflation are not modified by the isocurvature perturbations. The spectrum does not have bumps and is similar to that found in~\cite{Polarski:1995zn} in the case of two-field inflation driven by two massive inflaton fields with the standard kinetic terms and in the absence of a pronounced intermediate power-law stage between the two periods of inflation.

	\section{Primordial Black Hole formation}\label{sec:PBH}
\begin{figure}
		\centering
		\includegraphics[width=\columnwidth]{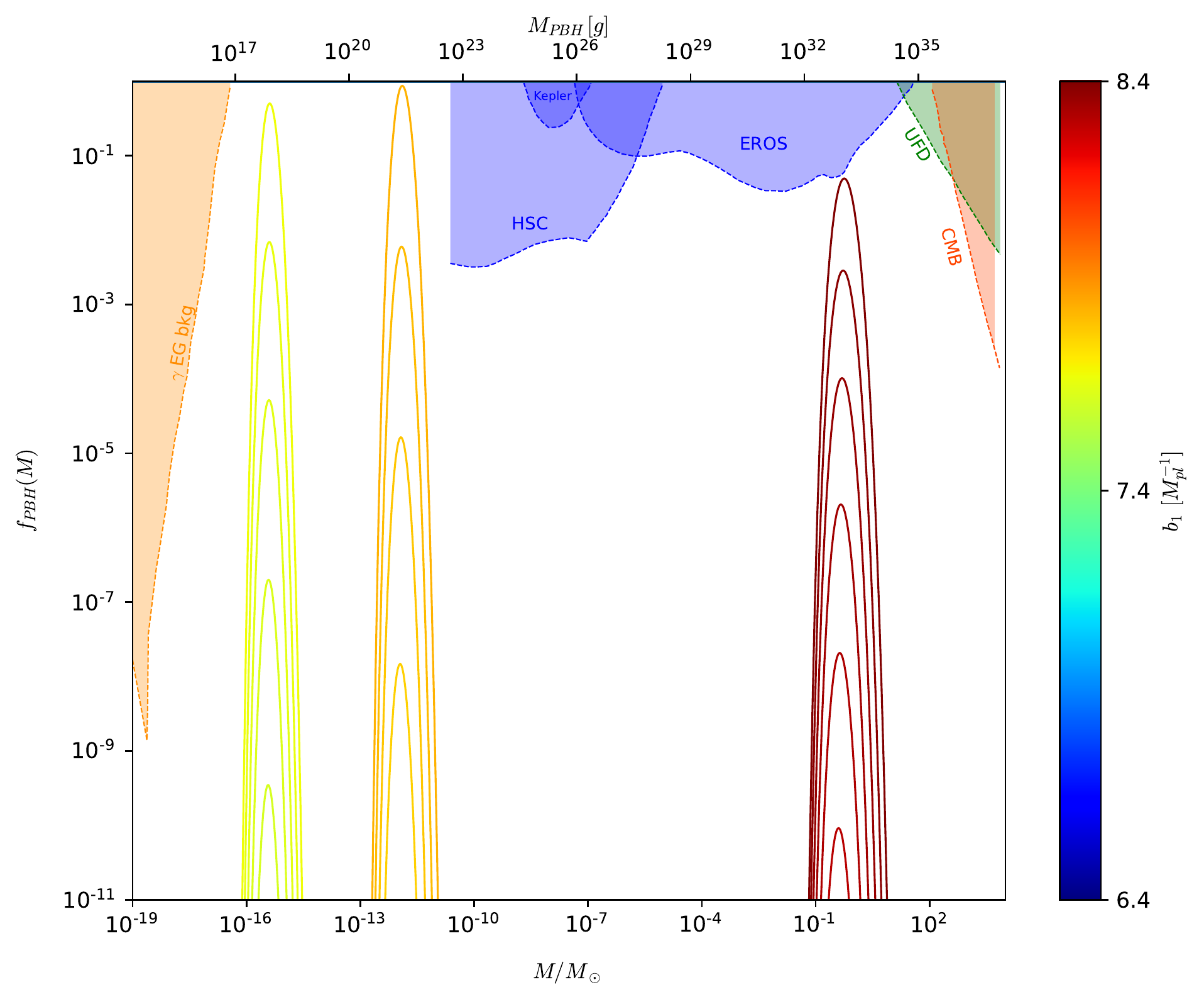}
		\caption{Fraction of PBH $f_\textup{PBH}$ as a function of the mass of the formed PBHs in solar masses and [g] units, computed from the spectra in~\autoref{fig:power3}. $b_1$ is varied for the same continuous range of values. The observational constraints represent those from extra-galactic radiation (EG bkg) \cite{Carr:2009jm},   microlensing by the Subaru  Hyper Suprime-Cam (HSC) \cite{Niikura:2017zjd}, Kepler \cite{Griest:2013esa}, EROS \cite{Tisserand:2006zx}, survival of ultra-faint dwarf galaxies (UFD) \cite{Brandt:2016aco} and the accretion on the CMB \cite{Horowitz:2016lib,Poulin:2017bwe,Blum:2016cjs,Ali-Haimoud:2016mbv}. We do not show constraints from long-livedness of white dwarfs (WD) \cite{Graham:2015apa},  the presence of neutron stars in globular clusters around $M\sim10^{-13} M_\odot$ \cite{Capela:2013yf} and femtolensing around $M\sim10^{-15} M_\odot$ \cite{Ricotti:2007au} since they have been contested in the literature, see e.g.  Refs.~\cite{Katz:2018zrn} and \cite{Montero-Camacho:2019jte}. Moreover, we do not show constraints from the 511 keV gamma-ray line from positrons in the Galactic center \cite{Laha:2019ssq,Dasgupta:2019cae}, which, although most stringent than the EG bkg, do not constrain our results further and constraints from 2nd order
gravitational waves~\cite{Chen:2019xse}, since they depend on additional
assumptions, see \cite{Carr:2020gox}. Note that all these constraints are constantly updated and improved. }
		\label{fig:fraction}
	\end{figure}

	 As mentioned above, if an overdensity in the early Universe is large enough, it  can collapse  to form a PBH when it re-enters the Hubble radius  during radiation dominated era~\cite{Carr:1974nx,GarciaBellido:1996qt}. A  useful parameter to investigate the PBH abundance is the mass fraction at formation $\beta(M)$ defined (for a Gaussian distribution of primordial fluctuation) as~\cite{Sasaki:2018dmp}
	\begin{equation}
	\beta(M)=2\gamma\int_{\delta_{c}}^{\infty}\,\frac{d\delta}{\sqrt{2 \pi}\sigma}e^{-\nu^2/2}=2\gamma \text{Erfc}(\nu/\sqrt{2}),
	\end{equation}
	where $\delta_c$ is the threshold energy contrast perturbation to PBH formation,  $\nu\equiv\delta_c/\sigma$ and
	\begin{equation}
	\sigma^2(M)\equiv\int_{0}^{\infty}\,d\,\ln k\,W^2(kR)\frac{16}{81}(k R)^4\,\mathcal{P}_{\mathcal{R}}(k)
	\end{equation}
	is the variance of the density fluctuation $\delta$. Note that these results are obtained by assuming a linear relation between the curvature perturbation and the gauge invariant density perturbations (see \cite{Germani:2018jgr} for a discussion on the dependence of the PBHs abundance on the shape of the PPS in this context), which matches the results obtained with the full non-linear analysis when the power spectrum is very peaked \cite{Germani:2019zez}. 
	The parameter $\gamma$ is a correction factor that we set to be $\gamma=0.2$ (as suggested by simple analytic calculations~\cite{Carr:1974nx}). We assume a conservative value of $\delta_c=0.35$ \cite{Sasaki:2018dmp} in what follows and we shall comment on other values in the concluding section. 
	The total fraction of PBHs against CDM  is given by
	\begin{equation}
	\label{ftotPBH}
	f^{\textup{tot}}_{\textup{PBH}}\equiv \frac{\Omega_\textup{PBH}}{\Omega_{CDM}}=\int\,d\,\ln M\frac{d f_{\textup{PBH}}(M)}{d\ln M},
	\end{equation}
	where 
	\begin{equation}
	\frac{d f_{\textup{PBH}}(M)}{d\ln M}=\nu(M)^2\Biggl\lvert\frac{d \ln \nu(M)}{d\ln M}\Biggr\rvert \,f_\textup{PBH}(M)
	\end{equation}
	and~\cite{Sasaki:2018dmp}
	\begin{equation}
	f_\textup{PBH}(M)=2.7\times10^8\left(\frac{0.2}{\gamma}\frac{M}{M_\odot}\sqrt{\frac{g_{*,f}}{10.75}}\right)^{-1/2}\beta(M),
	\end{equation} where $g_{*,f}$ is the number of relativistic degrees of freedom,
		is the fraction of PBHs against CDM at a given mass scale today. Given the mass of the formed PBH, it can be translated into a comoving scale $k$ using the relation: 
	\begin{equation}
	\frac{M(k)}{M_\odot}=30 \left(\frac{\gamma}{0.2}\right)\left(\frac{g_{*,f}}{10.75}\right)^{-1/6}\left(\frac{k}{2.9\times 10^5\text{Mpc}^{-1}}\right)^{-2}.
	\end{equation}
	Thus, using this formula we can estimate that the  power spectra in~\autoref{fig:power3}, which have a peak around $k_{\textup{SKA}}\sim 2\times 10^{6}$ $\text{Mpc}^{-1}$, $k_{\textup{LISA}}\sim  10^{12}$ $\text{Mpc}^{-1}$, $k_\textup{BBO}\sim8\times10^{13}$ $\text{Mpc}^{-1}$ and $k_{\textup{ET}}\sim6\times 10^{15}$  $\text{Mpc}^{-1}$ correspond to a  $f(M)$ peaked at $M_\textup{SKA}\sim 35 M_\odot$, $M_\textup{LISA}\sim \times 10^{-12} M_\odot$, $M_{\textup{BBO}}\sim3\times10^{-16}  M_\odot$ and $M_\textup{ET}\sim7\times 10^{-20} M_\odot$ respectively.  We plot $f_\textup{PBH}$ for the four examples in~\autoref{fig:fraction} together with  cosmological and astrophysical constraints~\cite{Inomata:2017okj}. Note that the constraints in~\autoref{fig:fraction} are derived assuming a monochromatic distribution of PBHs, whereas $f_\textup{PBH}$, though very narrow, is extended over a small range of a masses. However, a discussion of how our results are affected by constraints for broad mass functions (see e.g. Refs.~\cite{Carr:2017jsz,Carr:2020gox,Kalaja:2019uju}) is not the purpose of this paper.
	As can be seen from the colorbar in~\autoref{fig:fraction}, we can tune $b_1$ and $\chi_i$  to obtain the maximum fraction $f^{\textup{tot}}_{\textup{PBH}}$ allowed by the constraints. In particular we obtain $f^{\textup{tot}}_{\textup{SKA}}=0.01$ and  $f^{\textup{tot}}_{\textup{LISA}}=f^{\textup{tot}}_{\textup{BBO}}=1$.  PBHs lighter than $10^{-19} M_\odot$ would have already evaporated by today and thus we had to tune our parameters, so that $f^{\textup{tot}}_{\textup{ET}}\simeq0$. We note that $f^{\textup{tot}}_{\textup{SKA}}$ is in agreement with the limit arrived at by Refs.~\cite{Bird:2016dcv,Sasaki:2016jop,Clesse:2016vqa} though the spectral index of this configuration is  significantly lower than the present 95\% bounds from Planck.

	\section{Generating Gravitational Waves at small scales}\label{sec:GW}
	We now investigate the consequences of the bumps in~\autoref{fig:power3} concerning the  production of SBGW. Indeed, large scalar overdensities behaves as a (second order) source for a SBGW through second order perturbations~\cite{Acquaviva:2002ud,Inomata:2016rbd}.
	The energy density of the gravitational waves is given by~\cite{Baumann:2007zm,Espinosa:2018eve}:
	\begin{align}
	\label{eq:GW}
	\Omega_{\textup{GW}}=&\frac{\Omega_{r,0}}{36}\int_{0}^{\frac{1}{\sqrt{3}}}d\textup{d}\int_{\frac{1}{\sqrt{3}}}^{\infty}ds\,\left[\frac{(d^2-1/3)(s^2-1/3)}{s^2-d^2}\right]^2\notag\\
	&\cdot\mathcal{P}_{\mathcal{R}}\left(\frac{k\sqrt{3}}{2}(s+d)\right)
	\mathcal{P}_{\mathcal{R}}\left(\frac{k\sqrt{3}}{2}(s-d)\right)[\mathcal{I}_c(d,s)^2+\mathcal{I}_s(d,s)^2],\end{align}
	where $\Omega_{r,0}\simeq 8.6\times10^{-5}$ is the density of radiation today and the functions $\mathcal{I}_{c,s}$ are given in Eqs.~(D.1) and (D.2) of Ref.~\cite{Espinosa:2018eve}.
	
	\begin{figure*}
		\begin{center} 
				\resizebox{453pt}{170pt}{\includegraphics{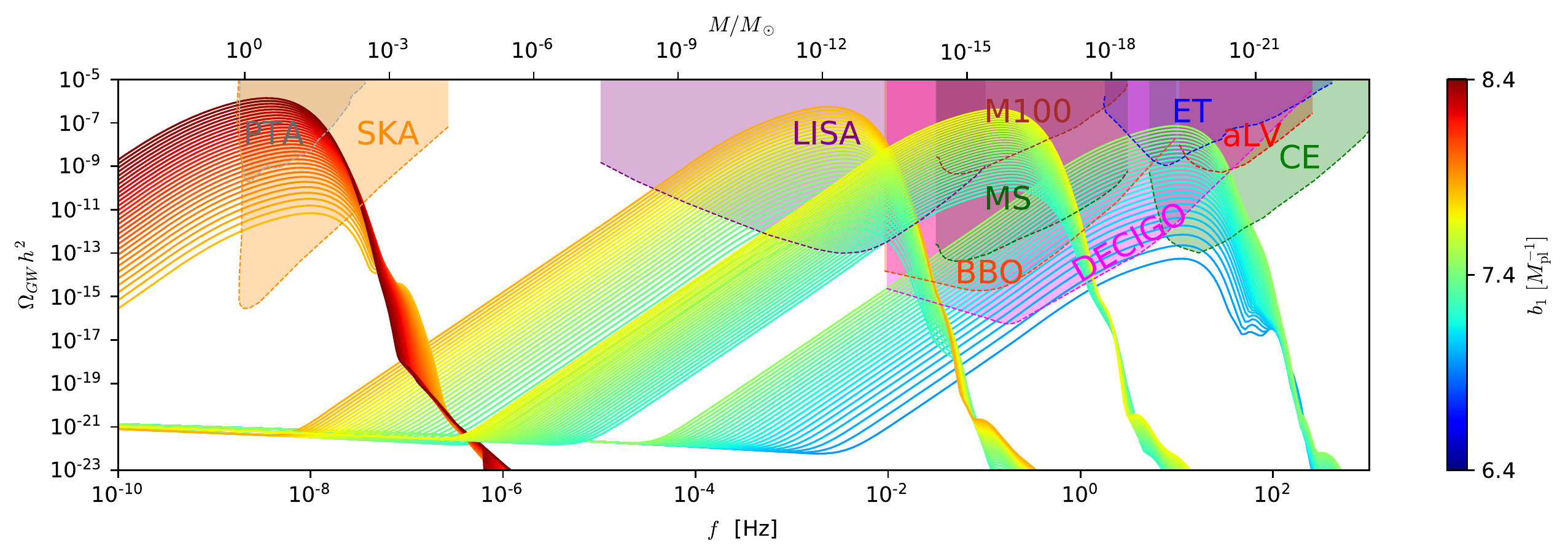}}
		\end{center}
		\caption{\label{fig:GW} Relic energy density  of gravitational waves  computed from the spectra in~\autoref{fig:power3}. $b_1$ is varied for the same continuous range of values.  }  
	\end{figure*}
In~\autoref{fig:GW}, we plot $\Omega_{\textup{GW}}h^2$, where we assume $h^2=0.49$, computed using~\autoref{eq:GW} and the power spectra in~\autoref{fig:power3} together with the sensitivity of the various forthcoming GW experiments. We plot four families of GW density corresponding to four frequency windows of future observations. In each family the variations in the peak height and position of the density are controlled by $b1$ shown in the colorbar.
	It is evident that the PTA limits on SBGW~\cite{Lentati:2015qwp,Shannon:2015ect,Aggarwal:2018mgp} already exclude some of the lines. This means that it is not possible  to produce the $f^{\textup{tot}}_{\textup{SKA}}=0.01$ quoted in the last section within the framework of our model. Nevertheless, there exist some values of the coupling $b_1$ for which $\Omega_\textup{GW}\,h^2$ falls within the sensitivity of the PTA search with SKA~\cite{Zhao:2013bba}. Moving to higher frequencies
	$\Omega_{\textup{GW}} \,h^2$ peaks well inside the range of detectability of LISA ~\cite{Audley:2017drz,Caprini:2015zlo} and  its tail can also be detected by DECIGO/BBO~\cite{Yagi:2011wg} for certain values of $b_1$. Also, in  the $BBO$ case, it peaks in the frequency range targeted by BBO and DECIGO. We stress that in the LISA and BBO cases, as can be seen from~\autoref{fig:fraction}, the large scalar density perturbations that source the stochastic background of GWs can also be responsible for the seeding of PBHs of $\sim10^{-12} M_\odot$ and $\sim10^{-16} M_\odot$ PBHs respectively, that can constitute up to the totality of the observed CDM in our Universe. A detection of such a signal from future space based intereferometers would be a strong hint of this possibility~\cite{Bartolo:2018evs,Bartolo:2018rku}.
	Finally, in the ET case, we show how our mechanism can operate even at higher frequencies. We have already taken care in~\autoref{sec:PBH} that the parameter chosen do not lead to PBH production as PBHs with those masses would have already evaporated today, leaving traces in the extragalactic $\gamma$ ray background~\cite{Carr:2009jm}.  In this range we found the striking result that $\Omega_{\textup{GW}} \,h^2$ (at least for the highest values of $b_1$) can be detected simultaneously by the DECIGO/BBO, Magis-AION-space~\cite{Coleman:2018ozp}, the Einstein Telescope~\cite{Sathyaprakash:2009xs}, Advanced Ligo + Virgo~\cite{TheLIGOScientific:2016dpb} and CE~\cite{Evans:2016mbw}.   
	
	We stress again that the height of the peak in $\Omega_\mathrm{GW}$ is only logarithmically dependent on $f^{\textup{tot}}_{\textup{PBH}}$ and thus detectable GWs can be produced even when there are no PBHs~\cite{Orlofsky:2016vbd}. 
	For this reason, the production of small scales  GWs is even more robust phenomenological prediction than PBH for our model.

\section{Changing the non-canonical coupling}\label{sec:coupling}
	In this section, we explore the sensitivity of the results of the previous sections to the functional form of the coupling. We assume $f_B(\phi)=\exp(2 b_2 \phi^2)$ as in~\autoref{eq:coupling2} and, to facilitate the comparison, we restrict to the LISA case. Besides the different form of $b(\phi)$ and $b_{\phi}$, the main difference between this coupling and $f_A(\phi)$ is a non-vanishing second derivative $b_{\phi\phi}=b_2=\rm{const}$.
	A non-vanishing $b_{\phi\phi}$  adds a new contribution to the change of curvature perturbation $\mathcal{R}$ and also modifies the effective mass of the isocurvature perturbations in~\autoref{eq:effmassiso}.  

\begin{figure*}
    	\begin{center} 
    		\resizebox{143pt}{115pt}{\includegraphics{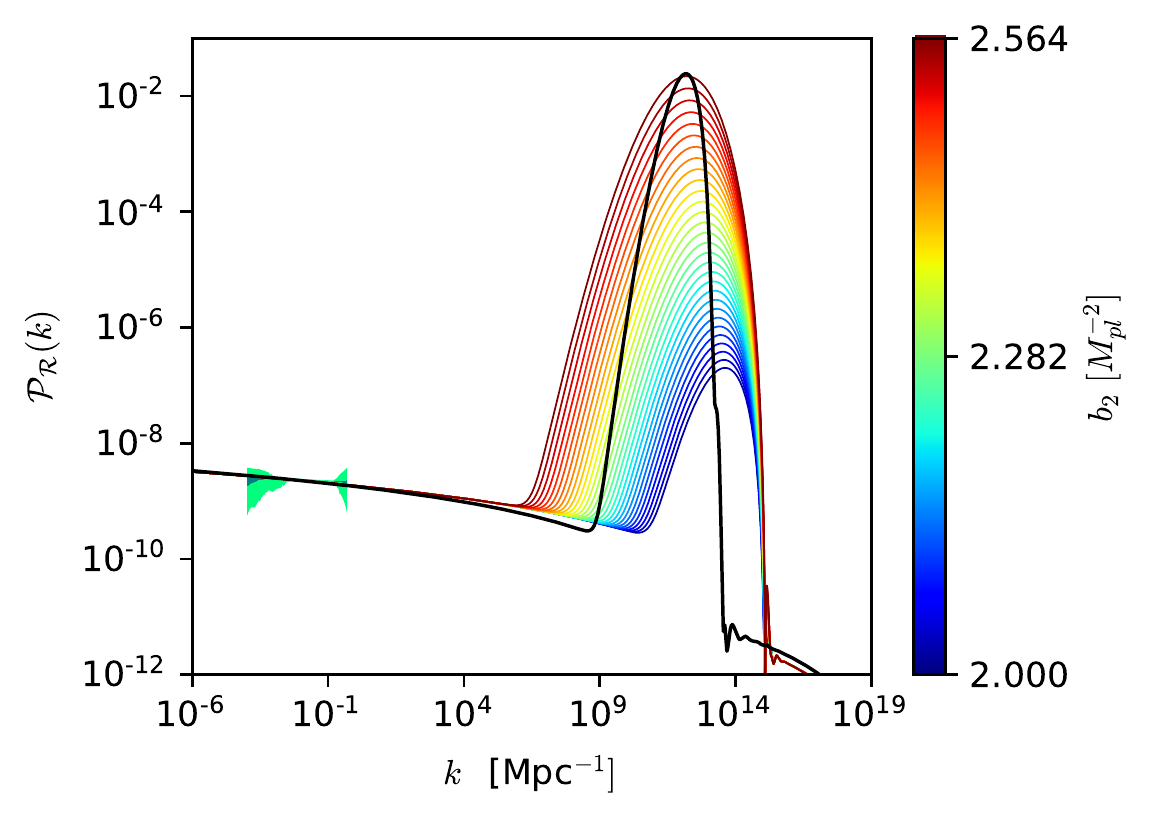}}
    		\resizebox{143pt}{120pt}{\includegraphics{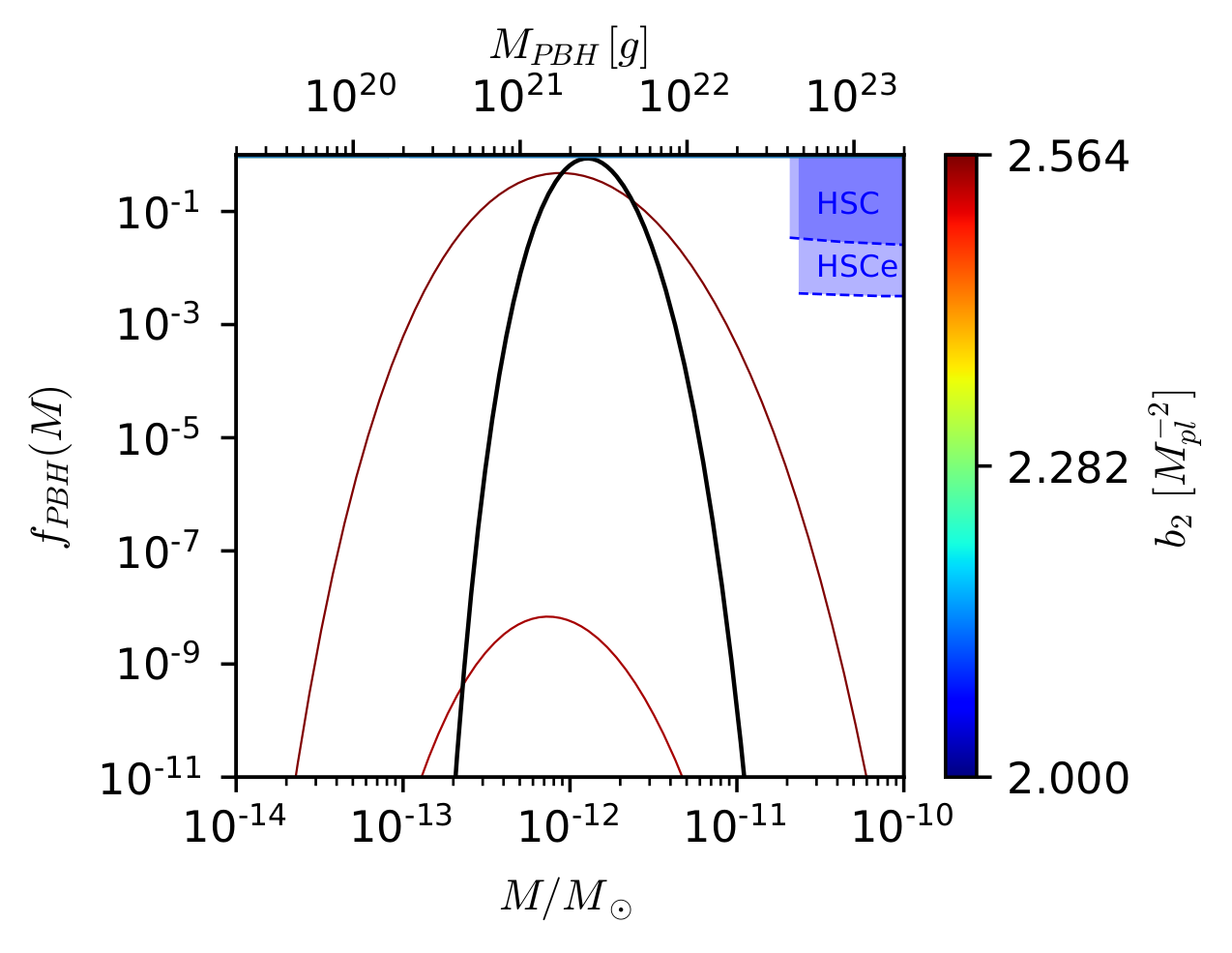}}
    		\resizebox{143pt}{115pt}{\includegraphics{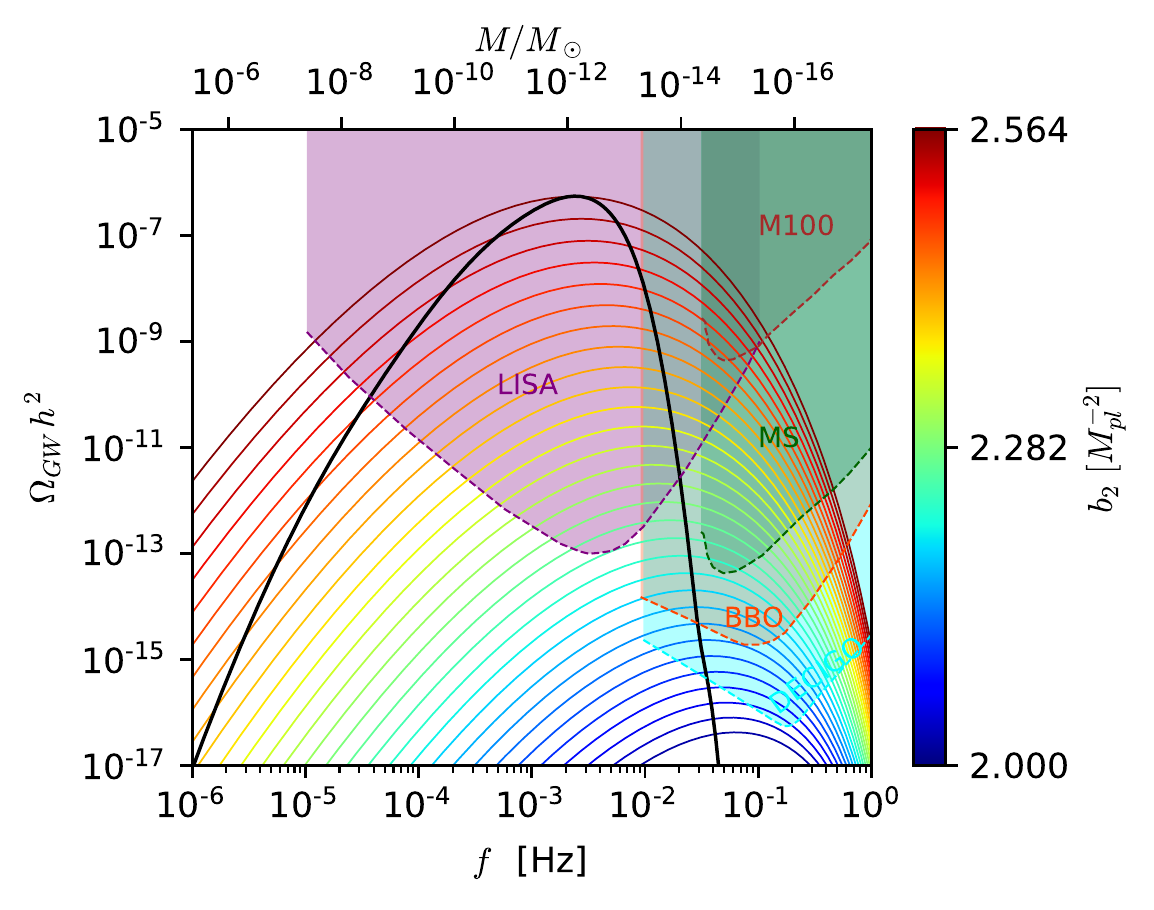}}
    	\end{center}
    	\caption{\label{fig:ResultsTwoCouplings} [Left] Scalar power spectra [Center] PBHs mass function and [Right] relic energy density of GWs for the $f_B$ model. We plot the results for the correspondent case in the $f_A$ model with $b_1=8.4 M\textup{pl}^{-1}$ in solid black lines. $b_2$ is varied for a continuous range of values. }    
    \end{figure*}

    In~\autoref{fig:ResultsTwoCouplings}, we show the results for the scalar power spectrum, PBHs mass fraction and induced SBGW for a range of values of the non canonical coupling $b_2$. As stated above, we have used the same parameters as the LISA case except for the initial conditions on the second inflaton $\chi_i$ that we have fixed to the lower value $\chi_i=6.8\,M_\textup{pl} $, in order for the peak in $\mathcal{P}_\mathcal{R}(k)$ to be at the same scale in the $f_A$ and $f_B$ case. Indeed, although the background evolution is essentially the same in the two cases, a non vanishing $b_{\phi\phi}$ makes the isocurvature tachyonic instability more prominent during the transition between the two stages of inflation. As a result a broader range of scales feel the isocurvature feedback and the peak has a broader structure. This affects both the large and small scale phenomenology of the model. At CMB scales, a smaller $\chi_i$ reduces the duration of the second stage of inflation and the spectral index is now given by $n_s=0.9628$, which is no more in tension with the CMB constraints.
    On the other hand, such a broad peak in the power spectrum modifies the mass fraction of primordial black holes. This is important as a broader mass function that extends to a larger mass range can lead to the totality of CDM in form of PBHs, that is $f_\textup{PBH}^\textup{TOT}=1$, even with a smaller peak.
    Finally, we see from the right panel in~\autoref{fig:ResultsTwoCouplings} that also $\Omega_\mathrm{GW}$ extends to a broad range of frequencies and, for some values of the coupling, falls in the sensitivity of several future experiments at the same time. The different shape of the GWs relic density $\Omega_\mathrm{GW}$ can therefore be used  to confirm or reject this scenario (or even tell the difference between the two non-canonical coupling) by reconstructing the GW signal in the lucky event of a detection of a SBGW by future GW experiments~\cite{Kuroyanagi:2018csn,Caprini:2019pxz}.

	\section{Discussion}
	\label{sec:discussion}

 We have presented a generic mechanism that operates in  models where inflation consists of two stages. The first stage of inflation is driven by an effectively heavier scalar field that eventually settles in its minimum and the second stage is driven by a non-canonically coupled lighter one. For large enough non-canonical coupling, a temporary tachyonic instability of the isocurvature perturbation feedbacks on the curvature perturbation and sources a large bump  in the primordial power-spectra, with a maximal rate of growth of $n_s=4$. If the bump is larger than a certain threshold, such  large perturbations collapse into PBHs when re-enter the Hubble radius during radiation dominated era. Furthermore, these large scalar fluctuations can source a SBGW to second order in perturbation theory. We have developed an extension of BINGO where we numerically solve for the primordial perturbations induced by two field inflationary scenario in a non-canonical Lagrangian. Using the obtained primordial spectrum we have computed the predicted mass fraction of the PBHs and and relic energy density of GWs for a coupling of the form $e^{2b_1\phi}$ in the case of four configurations representative for SKA, LISA, BBO and ET according to the frequency at which the resulting $\Omega_\mathrm{GW}$ peaks. We have shown that PBHs can be a significant fraction of CDM in the LISA and BBO case, although the former case is in slight tension with the CMB observations. On the other hand, the PBHs abundance is exponentially sensitive to the amplitude of $\mathcal{P}_\mathcal{R}(k)$ and even a small decrease in its amplitude can lead to a significantly smaller PBHs abundance, still producing a detectable SBGW, which is possible in all the four cases. 

 We have also analyzed the dependence of our results on the functional form of the coupling in the non-canonical kinetic term in the specific LISA case. We have found that for  a coupling of the form $e^{2b_2\phi^2}$, the isocurvature feedback to curvature perturbation is more efficient and a broader peak is produced, with important consequences on the resulting PBHs mass function and SBGW. Furthermore, for the broad bump to peak at the same frequency of the first coupling case, the second stage has to last shorter, reconciling the CMB predictions of the LISA case with observations.

Even though we assumed a particular  model for the potential, our mechanism is generic and works with every potential provided that it has an effective minimum for the heavy field to settle in. In particular our results on $n_s$ show that models with slightly bluer spectra are preferred as the peak shifts towards larger scales. 
It would be interesting to study a realistic model that naturally predicts our mechanism from a theoretical rather than a phenomenological point of view.

Finally, we have assumed a Gaussian statistics of primordial scalar fluctuations. It is well known that non-Gaussianity strongly affect the primordial abundance of PBHs~\cite{Young:2013oia,Garcia-Bellido:2017aan,Franciolini:2018vbk,Atal:2018neu,DeLuca:2019qsy,Yoo:2019pma,Ezquiaga:2019ftu} and the SBGW~\cite{Cai:2018dig}. It  would thus be of extreme importance to extend the findings of this paper to the computation of  the non-Gaussianities generated by our two-field model~\cite{Garcia-Saenz:2019njm}.

	\section*{Acknowledgements}
MB thanks Guillermo Ballesteros for useful discussions on PBH constraints. DKH has received fundings from the European Union’s Horizon 2020 research and innovation programme under the Marie Sklodowska-Curie grant agreement  No. 664931.
FF acknowledges financial 
support by ASI Grant 2016-24-H.0.
LS wishes to acknowledge support from the Science and Engineering Research  
Board, Department of Science and Technology, Government of India, through 
the Core Research Grant CRG/2018/002200. AAS was partially supported by the Russian Foundation for Basic Research grant No. 20-02-00411.

\vspace{1cm}	
	
\noindent	
\textbf{Note added:}
While this project was nearly complete, two related papers \cite{Palma:2020ejf,Fumagalli:2020adf}, also studying the production of PBHs from turns in the field space, appeared on the arXiv. These works discuss the enhancement of the power spectrum due to turning trajectories in multi-field inflation and, in addition to numerical results, provide approximate analytical  solutions  assuming  a top hat \cite{Palma:2020ejf} and Gaussian profile \cite{Fumagalli:2020adf} for $\dot{\theta}$ that are found to be adequate for a sharp and smooth gradual turn respectively. 
Differently from those papers, we focus here on a concrete example of an inflationary model consisting of two stages. As can be seen from~\autoref{fig:Background}, the oscillations of the  field induce an oscillatory pattern for $\dot{\theta}$ for which the analytical results of \cite{Palma:2020ejf,Fumagalli:2020adf} are only qualitatively applicable and only a numerical integration can give accurate results. In addition to that, we have computed the GW energy density at all scales relevant to future observations, which is one of the key results of this paper.

\appendix


\section{Varying the ratio of the potentials}\label{appendix:VaryingMass}	
In this Appendix, we analyze the effect of changing the potential ratio $R\equiv V_0/(m_\chi\,M_\textup{pl})^2$ in our model. As in~\autoref{sec:coupling}, we focus on the LISA case. We vary the parameters according to Table \ref{tab2},
	\begin{table*}\begin{tabular}{ |p{2cm}||p{2.2cm}|p{2.2cm} |p{2.2cm}   |p{2.2cm}    |}
	\hline
	&  $\chi_i \,[M_\textup{pl}]$&$ V_0\,[10^{-10} M_\textup{pl}^4]$ &$R$&$b_1 [M_\textup{pl}]^{-1}$ \\
	\hline
	1  & 3.2& 6.4 &30 &9.466\\
	2&  7.31 &7.08  &500&7.837 \\ 
	3 & 8.1 &7.6& 1050&  7.382\\
	4& 8.5 &8.21& 3800&6.233 \\
	\hline
\end{tabular}\caption{\label{tab2} 
	Parameters used to reproduce Fig.~\ref{fig:ResultsPotentialRatio}.}
\end{table*}
\bigbreak
and keep the initial condition on the inflaton driving the first inflationary stage fixed to the value used in the main text. We have chosen the parameters to get $f^\textup{TOT}_\textup{PBH}\simeq 1$ in all the cases considered.

\begin{figure}
    	\begin{center} 
    		\resizebox{214pt}{172pt}{\includegraphics{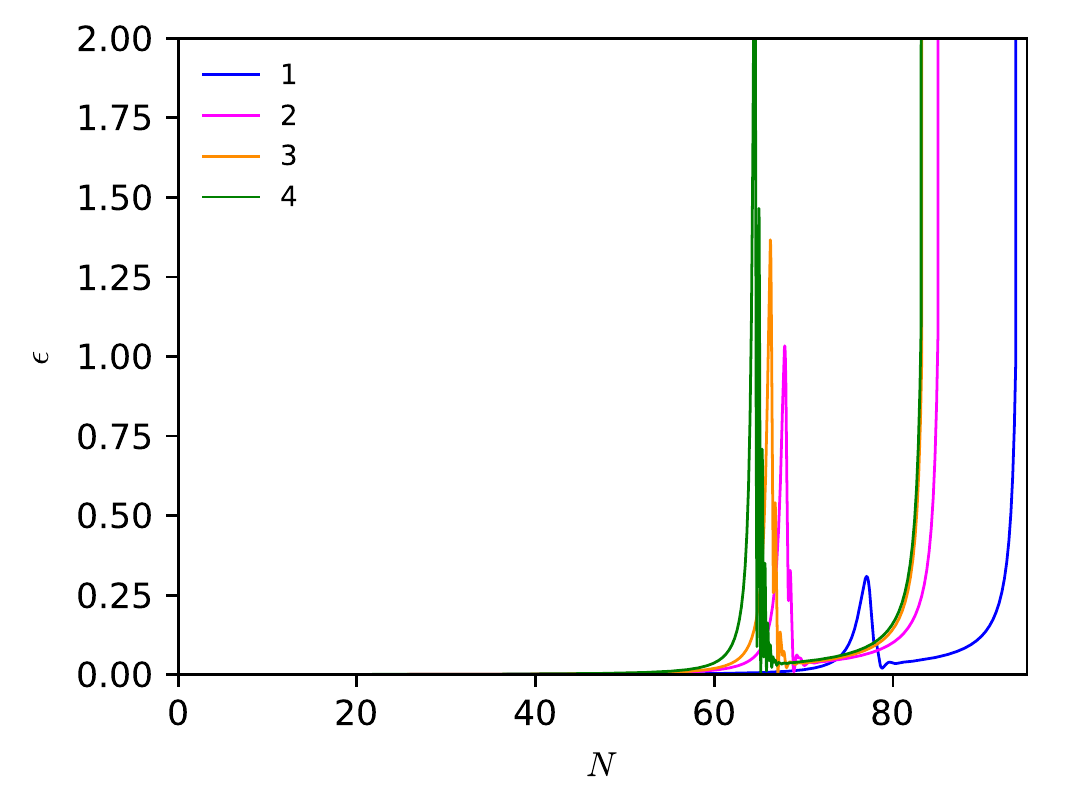}}
    		\resizebox{214pt}{172pt}{\includegraphics{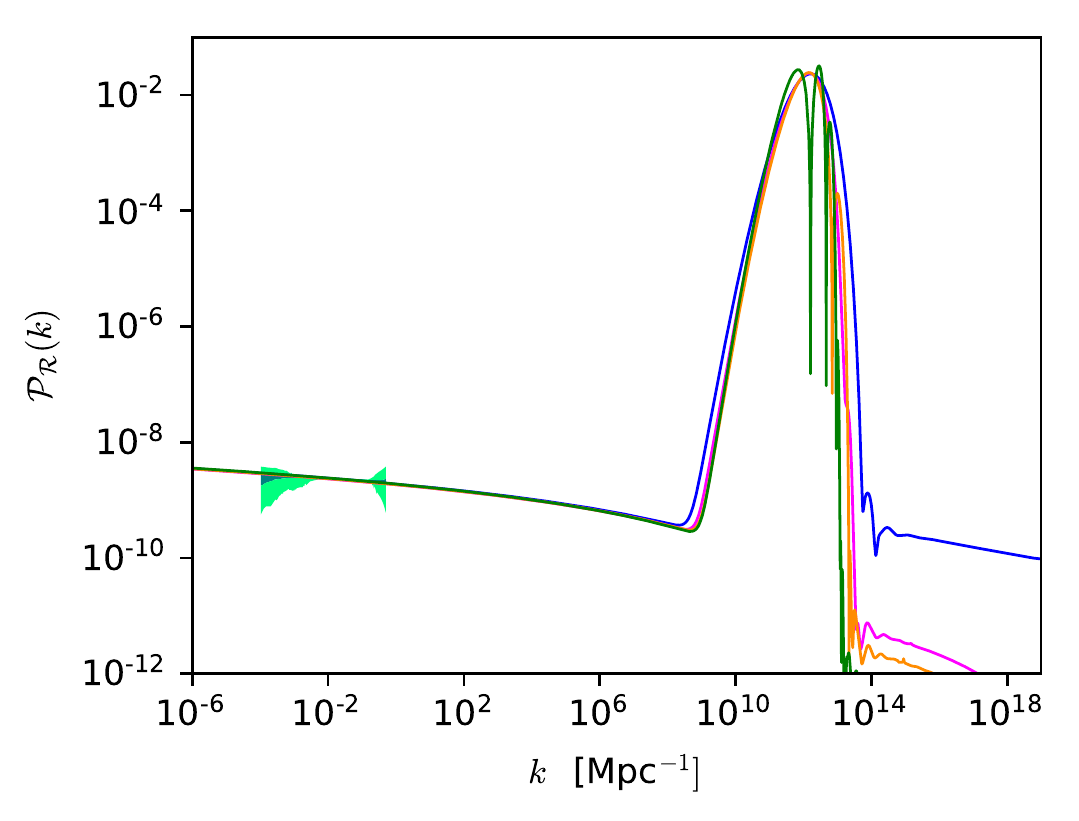}}
    		\resizebox{214pt}{172pt}{\includegraphics{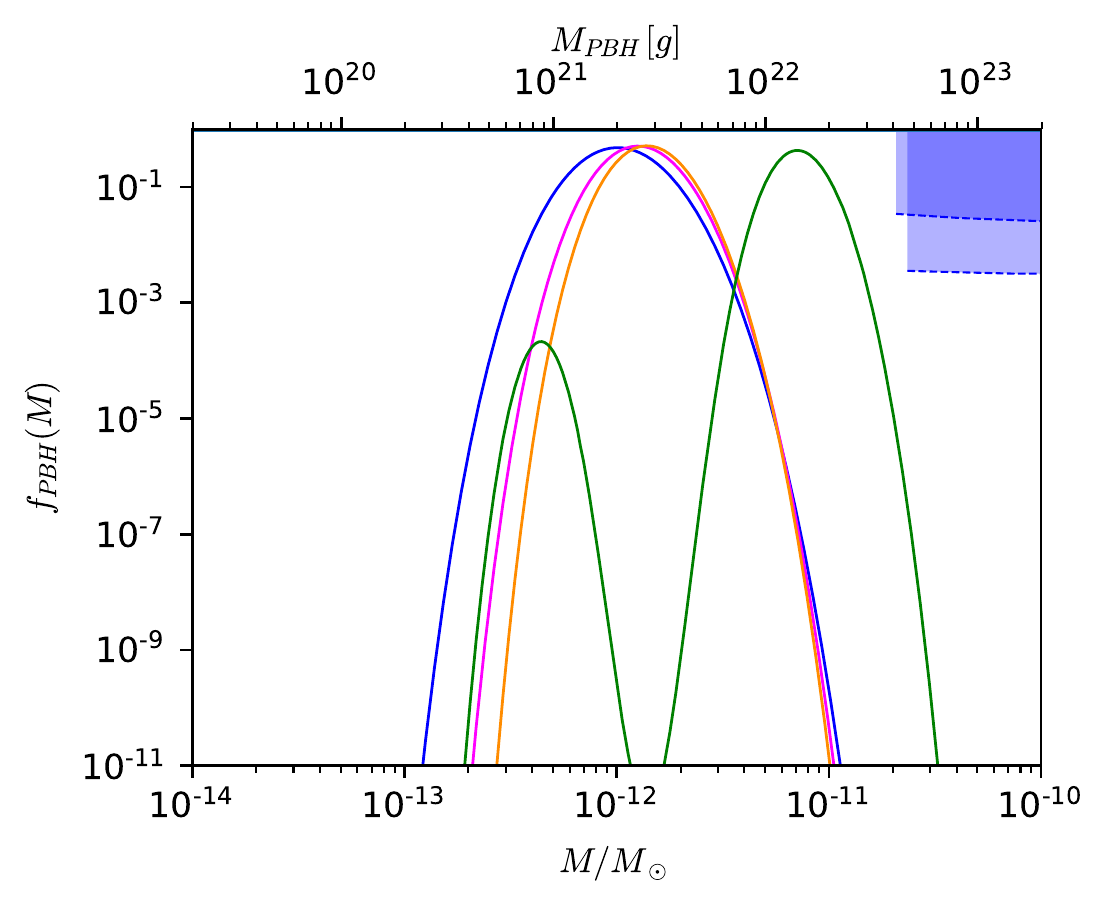}}
    		\resizebox{214pt}{172pt}{\includegraphics{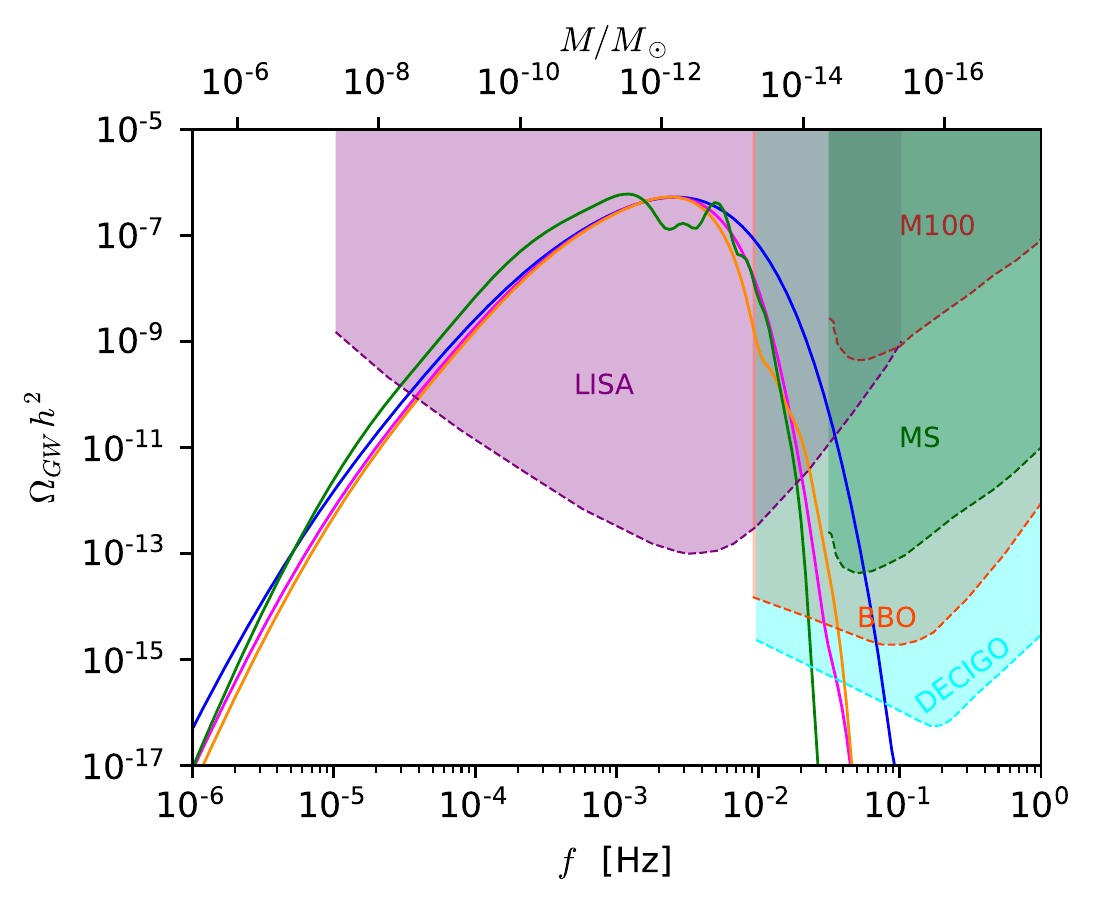}}
    	\end{center}
    	\caption{\label{fig:ResultsPotentialRatio} [Top-left] $\epsilon$ parameter, [top-right] Scalar power spectra, [bottom-left] PBHs mass function and [bottom-right] relic energy density of GWs for the $f_B$ model. We plot the results for the correspondent case in the $f_A$ model with Using the values in the Table in the main text. }    
    \end{figure}

We show our results in~\autoref{fig:ResultsPotentialRatio}. As can be seen, the slow-roll violation gets more violent when the ratio between the two potential is higher. In fact, for the case 3 and 4, $\epsilon$ becomes larger than $1$ and inflation ends at the transition to start again driven by the second lighter scalar field. This resembles a phase of intermediate matter-domination that is well known to occur in the case of two massive inflaton when the mass ratio is large enough 
	~\cite{Polarski:1992dq,Polarski:1994rz}. 
	
	The different pattern of the slow-roll violation is clearly imprinted in the scalar power spectrum. For larger values of the potential ratio $R$, in fact, we note an oscillatory the bump splits in a series of different peaks. This multi-peaked shape modifies the PBHs mass function, that becomes sharper as $R$ increases. In the extremal   case 4, the first two peaks in the scalar power spectrum have a comparable amplitude and give rise to an interesting PBHs mass function with a larger peak around $M\sim 10^{-10} M_\odot$ and
    a smaller one around $M\sim 3\times 10^{-12} M_\odot$.
    
    Furthermore, also the spectral shape of the relic GWs energy density is very different in the four cases considered. We stress again the phenomenological importance of predicting  distinct different shapes for $\Omega_\mathrm{GW}$ view of the signal reconstruction program with future GWs experiments~\cite{Caprini:2019pxz}.

		\section{Background equations and analytical results}
	\label{appendix:Background}
	In this Appendix, we review the basic  background equations and collect some useful analytical results that are valid before and after the transition between the first and second stage of inflation.

	From \autoref{action}, the equations of motion governing 
	the homogeneous scalar fields and the Friedmann equations are given by
	\begin{subequations}
		\label{eq:sf}
		\begin{eqnarray}
		\label{eq:KGphi}
		\ddot{\phi}+3 H\dot{\phi}+U_\phi
		&=&b_\phi e^{2 b}\dot{\chi}^2,\\
		\label{eq:KGchi}
		\ddot{\chi}+(3 H+ 2 b_\phi \dot{\phi}) \dot{\chi}
		+{\rm e}^{-2 b}W_\chi&=&0,
		\end{eqnarray}
	\end{subequations}
	\begin{subequations}
		\label{eq:f}
		\begin{eqnarray}
		H^2&=&\frac{1}{3 \Mpl^2}		\left[\frac{\dot{\phi}^2}{2}		+{\rm e}^{2 b}\frac{\dot{\chi}^2}{2}+V\right],\\
		\dot{H}&=&-\frac{1}{2 \Mpl^2}\left[\dot{\phi}^2
		+{\rm e}^{2b}\dot{\chi}^2\right].
		\end{eqnarray}
	\end{subequations}
	where, we write our potential \autoref{eq:potential} as $V(\phi,\,\chi)=U(\phi)+W(\chi)$.	
	When  $\phi$ and $\chi$ are slow-rolling, the equations above can be simplified by neglecting second time derivatives and products of squared first time derivatives and are approximated by (using $\doteq$ to denote an equality that is valid only assuming slow-roll for $\phi$ and $\chi$):
	\begin{subequations}
		\begin{eqnarray}
		\label{eq:KGphiSR}
		\dot{\phi}
		&\doteq&-\frac{U_\phi}{3 H},\\
		\label{eq:KGchiSR}
		\dot{\chi}&\doteq&-e^{-2 b(\phi)}\frac{W_\chi}{3H},\\
		\label{eq:hSR}
		H^2&\doteq&\frac{1}{3 \Mpl^2}		V,\\
		-\frac{\dot{H}}{H^2}&\equiv&\epsilon\doteq(\epsilon_\phi+\epsilon_\chi)
		\end{eqnarray}
	\end{subequations}
	where
	\begin{eqnarray}
	\epsilon_\phi&\equiv&\frac{M_\textup{pl}^2}{2}\left(\frac{U_\phi}{V}\right)^2\\
	\epsilon_\chi&\equiv&\frac{M_\textup{pl}^2}{2}\left(\frac{W_\chi}{V}\right)^2 e^{-2 b(\phi)}	.
	\end{eqnarray}

		\begin{figure}
		\begin{center} 
			\resizebox{214pt}{172pt}{\includegraphics{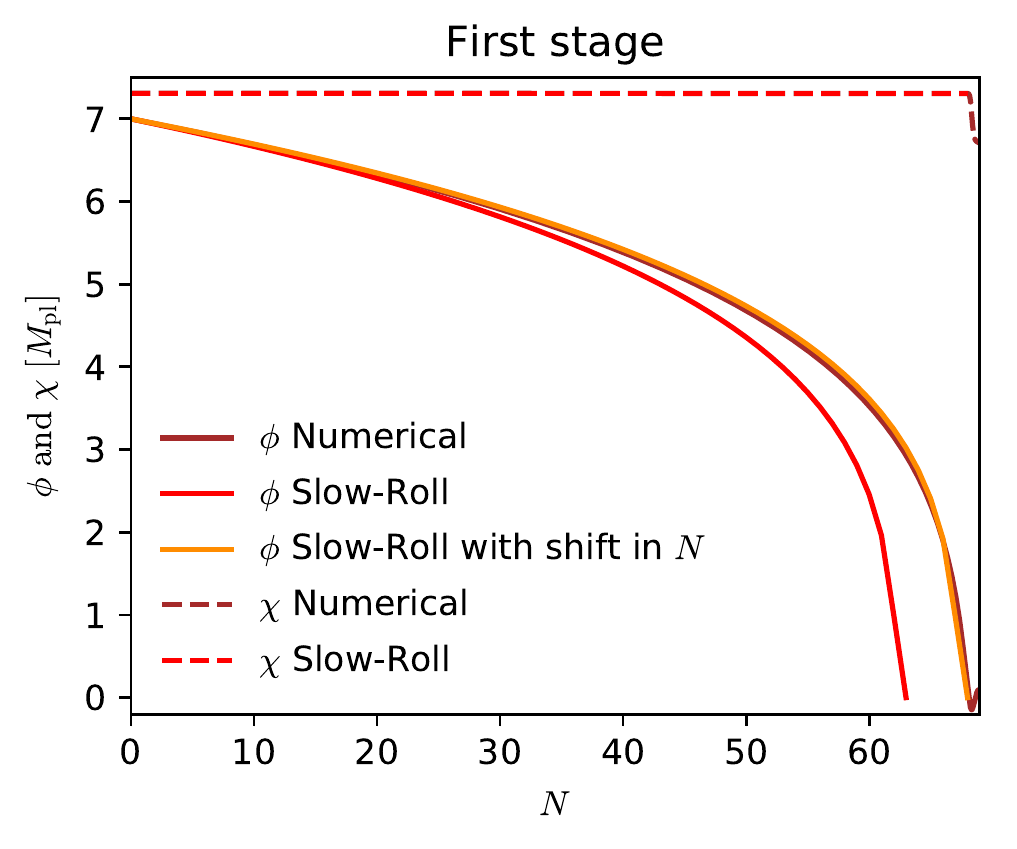}}
			\resizebox{214pt}{172pt}{\includegraphics{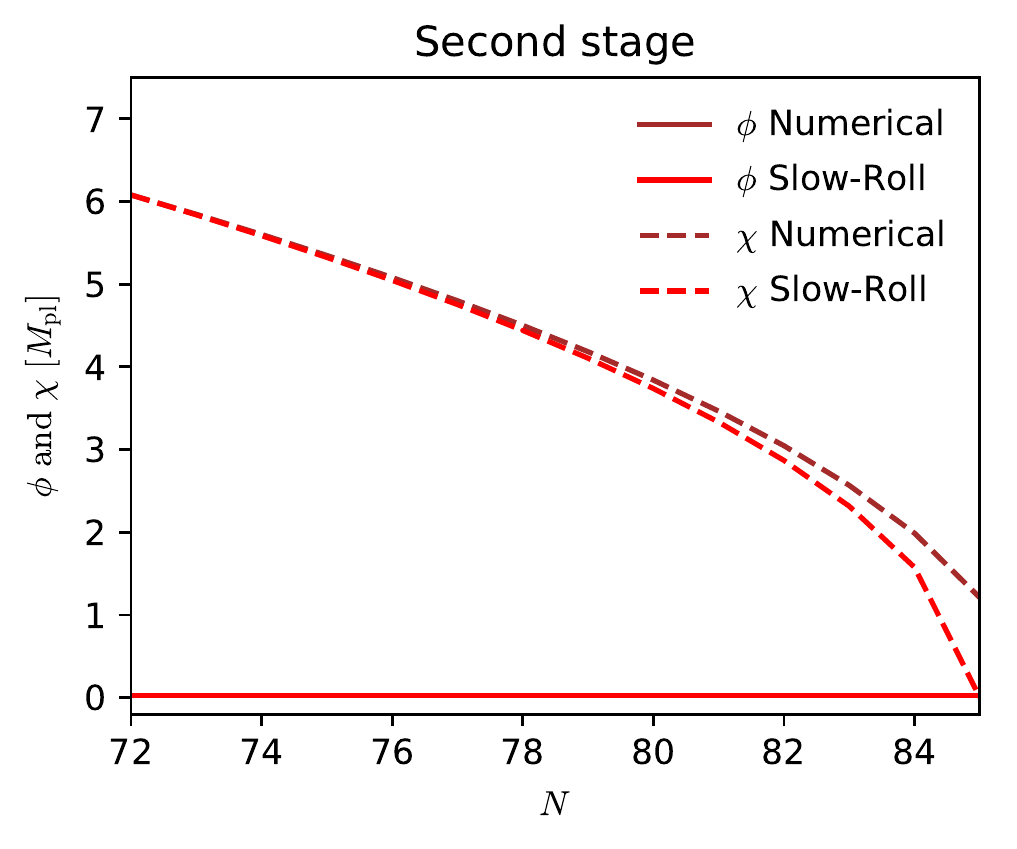}}
		\end{center}
		\caption{\label{fig:Comparison} [Left] Evolution of $\phi$ and $\chi$ during the first   and [Right] second  stage of inflation. We use the parameters for the LISA case with $b_1=7.3 M_\textup{pl}^{-1}$.}    
	\end{figure}

    During the first stage of inflation both $\phi$ and $\chi$ are slowly rolling.
    Using the slow-roll equation for $\phi$ \autoref{eq:KGphiSR}, we obtain:
    \begin{equation}
    \label{eq:phi1}
    \phi_1(N)=\sqrt{-\phi^2_0+\sqrt{-8 N\phi_0^4+\left(\phi_i^2+\phi_0^2\right)^2 }},
    \end{equation}
	where $\phi_i=\phi(0)$. We plot this solution in the left panel of \autoref{fig:Comparison}, in which we have considered the LISA case with $b_1=7.3 M_\textup{pl}^{-1}$ as an example. As can be seen, the analytical solution captures very well the numerical behavior. We note that the agreement can be make even better by shifting the argument in \autoref{eq:phi1} by a constant, which can be obtained by considering higher order slow-roll corrections to \autoref{eq:KGphiSR}.   
	
	During the first slow-roll stage we have $b_A(\phi)\gg1$ (or $b_B(\phi)\gg1$) and the term $\exp(-2b(\phi))$ may be taken as $0$ in \autoref{eq:KGchi}, leading to a constant $\chi_1(N)=\chi_i$, in perfect agreement with numerical results as shown in the left panel of \autoref{fig:Comparison}.
	
	The first and the second stage are separated by a transition which lasts, during which $\phi$ undergoes damped oscillations around the minimum, behaving as a massive scalar field, and $H$ and the field $\chi$ experience a jump as can be seen from the first panel in \autoref{fig:Background}, for which is not possible to obtain an analytical solution. Nevertheless, it is easy to obtain approximate expression after the decay of the oscillatory part of $\phi$, at, say, $N_2$. 
	
	To arrive at an expression for $\chi$ during this slow roll regime, we first note that the field $\phi$ approaches a constant value, given by the minimum of its effective potential, that we denote by $\phi_{\rm min}$. Denoting also the value of $\chi$ at the onset of this period as $\chi_{2,\,i}=\chi_1-\Delta\chi$, where $\Delta\chi$ is the jump in $\chi$, we can solve \autoref{eq:KGchiSR} and write:
	
	\begin{equation}
	\chi_2(N)=\sqrt{\chi_{2,\,i}^2 +4 M_\textup{pl}^2\left(N_2-N\right)e^{-2 b(\phi_2)}}. 
	\end{equation}
	We can then insert \autoref{eq:KGchiSR} in the equation of motion of the $\phi$, that is \autoref{eq:KGphi}, and the ansatz $\phi_2(N)=\phi_{\rm min}+\Delta\phi(N)$ to obtain:
	\begin{equation}
	\phi_{\rm min}= \begin{cases}
	\frac{b_1 m_\chi^2 \phi_0^2 M_\textup{pl}^2 }{3 V_0} &\text{for $b_A(\phi)$}\\
	0 &\text{for $b_B(\phi)$}.
	\end{cases}
	\end{equation}
    Note that, for the $B$ case, a second solution also exists, which is given by $\exp(2b_2\phi_{\rm min}^2)=2 b_2 m^2_{\chi}\phi_0^2M_{pl}^2/3V_0$ which, however, is not possible in our case since rhs of this equation is
less than unity for the numbers used in the paper.

	The equation for $\Delta\phi$ is instead given by
    \begin{equation}
	\Delta\phi''+\left(3+(\ln H)'\right)\Delta\phi'+\frac{m^2_{\Delta\phi}}{H^2}\Delta\phi=0
	\end{equation} 
	where a prime $'$ denotes a derivative with respect to the number of $e$-folds $N$. The effective mass square $m_{\Delta\phi}^2$ can be obtained by assuming $b_1 \phi \ll1$ $(b_2\phi^2\ll1)$ and linearizing $\exp[b(\phi_{\rm min}+\Delta\phi)]$ to arrive at 
	\begin{equation}
	m^2_{\Delta\phi}= \begin{cases}
	2\frac{V_0}{\phi_0^2}+\frac{4 b_1^2 m_\chi^2 \phi_0^2 M_\textup{pl}^2 }{3 } &\text{for $b_A(\phi)$}\\
	2\frac{V_0}{\phi_0^2}+\frac{4 b_2 m_\chi^2 \phi_0^2 M_\textup{pl}^2 }{3 } &\text{for $b_B(\phi)$}.
	\end{cases}
	\end{equation}
	Note that $m^2_{\Delta\phi}$ is always positive so that $\phi$ never experiences a tachyonic instability.
	
	The evolution of $\phi$ and $\chi$ after the oscillations have decayed is shown in the right panel of \autoref{fig:Comparison}, showing an overall good agreement.

	\bibliographystyle{JHEP}
	\bibliography{PBHNEW}
\end{document}